\documentclass[sigconf]{acmart}

\usepackage{amsmath,amssymb,amsfonts}
\usepackage{graphicx}
\usepackage{textcomp}
\usepackage{xcolor}
\usepackage{svg}
\usepackage{subfigure}
\usepackage{url}
\usepackage{tikz}
\usepackage{enumitem}
\usepackage{algorithm}
\usepackage{algpseudocode}
\usepackage{balance}
\usepackage{comment}
\AtBeginDocument{%
  }

\copyrightyear{2024}
\acmYear{2024}
\setcopyright{acmlicensed}\acmConference[MEMSYS '24]{The International Symposium on Memory Systems}{September 30-October 3, 2024}{Washington, DC, USA}
\acmBooktitle{The International Symposium on Memory Systems (MEMSYS '24), September 30-October 3, 2024, Washington, DC, USA}
\acmDOI{10.1145/3695794.3695805}
\acmISBN{979-8-4007-1091-9/24/09}

\begin{document}

\title{ZipCache: A DRAM/SSD Cache with Built-in Transparent Compression}


\author{Rui Xie}
\affiliation{%
 \institution{Rensselaer Polytechnic Institute}
 \city{Troy}
 \state{NY}
 \country{USA}}
\email{xier2@rpi.edu}

\author{Linsen Ma}
\affiliation{%
 \institution{Rensselaer Polytechnic Institute}
 \city{Troy}
 \state{NY}
 \country{USA}}
\email{mal3@rpi.edu}

\author{Alex Zhong}
\affiliation{%
  \institution{Harker School}
  \city{San Jose}
  \state{CA}
  \country{USA}}
\email{25alexz@students.harker.org}

\author{Feng Chen}
\affiliation{%
  \institution{Louisiana State University}
  \city{Baton Rouge}
  \state{LA}
  \country{USA}}
\email{fchen@csc.lsu.edu}

\author{Tong Zhang}
\affiliation{%
 \institution{Rensselaer Polytechnic Institute}
 \city{Troy}
 \state{NY}
 \country{USA}}
\email{tzhang@ecse.rpi.edu}

\renewcommand{\shortauthors}{Rui Xie, Linsen Ma, Alex Zhong, Feng Chen, and Tong Zhang}

\begin{abstract}
As a core component in modern data centers, key-value cache provides high-throughput and low-latency services for high-speed data processing. 
The effectiveness of a key-value cache relies on its ability of accommodating the needed data.
However, expanding the cache capacity is often more difficult than commonly expected because of many practical constraints, such as server costs, cooling issues, rack space, and even human resource expenses. 
A potential solution is {\it compression}, which virtually extends the cache capacity by condensing  data in cache. 
In practice, this seemingly simple idea has not gained much traction in key-value cache system design, 
due to several critical issues: the compression-unfriendly index structure, severe read/write amplification, wasteful decompression  operations, and heavy computing cost. This paper presents a hybrid DRAM-SSD cache design to realize a systematic integration of data compression in key-value cache. By treating compression as an essential component, we have redesigned the indexing structure, data management, and leveraged the emerging computational SSD hardware for collaborative optimizations. We have developed a prototype, called ZipCache. Our experimental results show that ZipCache can achieve up to 72.4\% higher throughput and 42.4\% lower latency, while reducing the write amplification by up to 26.2 times.  
\end{abstract}




\maketitle

\section{Introduction}\label{sec:introduction}
Key-value cache plays a crucial role in providing high-throughput, low-latency data services. Major Internet service providers, such as Google and Meta, often deploy a fleet of cache servers as the first line of defense to handle a massive influx of requests for key-value data, a type of unstructured data organized in simple forms as {\it keys} and {\it values}~(e.g., ``User ID'' and ``User name''). Key-value caching can accelerate data retrievals and alleviate the traffic to backend databases by serving from high-speed storage medium, typically DRAM, flash memory, or a combination of both, such as Meta's CacheLib~\cite{CacheLib-link} and Ximalaya's xcache~\cite{xcache-link}. 

The effectiveness of key-value caching hinges on its ability of accommodating the requested data in cache. While a larger cache intuitively leads to improved performance, expanding the capacity of key-value caches within data centers is often not simply a matter of hardware upgrade. Many practical constraints, such as server costs, cooling expenses, rack space, real estate limitations, and even human resource expenses must be taken into consideration. Let us consider the hardware cost as an example: Microsoft Azure reports that DRAM constitutes 50\% of their server costs~\cite{Morgan20}, and Meta reports a similar trend~(40\% of the rack cost)~\cite{Maruf_2023}; Although the prices of high-speed NVMe SSDs are comparatively lower, they are still rather substantial~\cite{White23, Olsan23, Ao23}. 
Relying solely on hardware investment to increase cache capacity is apparently not a sustainable, cost-effective approach to keep up with the rapid growth of data. This poses an increasingly severe challenge in today's data centers. 

A potential solution is {\it compression}. By condensing data to occupy a smaller footprint, one could {\it virtually} expand the cache capacity, allowing cache to hold more data, which in turn increases the cache hit ratio. Despite the adoption of data compression in computing systems in prior studies~\cite{choukse2018compresso,ekman2005robust,pekhimenko2013linearly,zhao2015buri}, interestingly, this simple idea has not gained much traction in key-value cache design. We believe that this lack of adoption in practice is due to several unique and critical issues inherent in key-value cache systems: 

\noindent $\bullet$ {\bf Issue \#1: The commonly used hash indexing causes random, compression-unfriendly data placement}. Most key-value cache systems adopt a {\it hash index} based structure to manage the key-value data~\cite{Redis-link, lim2014mica, 
chandramouli2018faster}. With hash indexing, keys are randomly dispersed in a flat, shallow data structure, which is advantageous for quick search in a large key space but comes with a detrimental effect for compression: Due to the  nature of hash functions, the keys are evenly distributed, leaving unrelated data randomly mingled together. Such data layout is inherently difficult for effective data compression, as compression algorithms heavily rely on organizing similar data content within a close proximity.

\noindent $\bullet$ {\bf Issue \#2: The structure designed for managing small-size key-value items introduces a severe read/write amplification problem}. Key-value workloads are known to be dominated by small-size data items. According to a study from Meta/Facebook, the majority of key-value items are (much) smaller than 500~bytes~\cite{Atikoglu12}. Since compressing each individual small key-value item yields limited or no benefits in size reduction, achieving effective compression requires to pack a collection of small key-value items  for a reasonable compression ratio. However, this ``optimization'' would result in a substantial, undesirable increase in access operations, i.e.,~{\it read/write amplification}, when reading or updating a small key-value item in a much larger compression unit.  

\noindent $\bullet$ {\bf Issue \#3: Compression and decompression are simply treated as two opposite processes on the same unit of data}. As a common practice, a block of data is compressed and decompressed as a full, single unit.  
As we increase the compression granularity to reduce indexing costs and increase compression ratios, the efficiency of decompression process unfortunately diminishes. This is because the more data is compressed, the more needs to be decompressed, leading to a proportionally increased amount of data accesses and longer delays to decompress and locate the requested key-value item. 

\noindent $\bullet$ {\bf Issue \#4: Compression imposes a heavy computing cost and interferes other data-processing tasks}. It is well-known that compression is computation intensive, essentially trading computation for storage capacity. Conducting data compression and decompression operations on general-purpose CPUs not only increases the burden on limited CPU resources but also causes disruptive effect on foreground service operations, potentially reducing the overall system throughput and increasing user-perceived delays. Considering the stringent requirement for cache latency, the additional delay is a non-trivial overhead that must be considered and mitigated. 


All the above-said issues pose a critical challenge to the current key-value cache systems, calling for a full consideration of data compression as an essential component in the cache system design. 
In this paper, we propose a new scheme, called {\it ZipCache}, to realize a systematic integration of data compression in the key-value cache system design. By treating data compression as an integral component, we have redesigned the indexing structure, data management, and leverage cutting-edge hardware for collaborative optimizations. 
Specifically, we take several important measures to achieve systematic optimizations for data compression: 

Firstly, we abandon the conventional hash indexing structure and adopt a seemingly more ``costly'' B+~tree based structure to manage key-value items. This enables us to preserve content similarity and retain the spatial locality. Secondly, we introduce a sparse structure, called {\it super-leaf}, to store key-value data for compression in a virtualized SSD storage space. This leverages the emerging commercial SSDs with built-in transparent compression to maintain low-cost indexing without wasting any physical storage space. Thirdly, we decouple the data units for compression and decompression by creating a special intra-page structure for {\it just-in-need decompression}. This method facilitates early termination, significantly reducing decompression time and read amplification. Lastly, we fully exploit the abilities of the emerging computational SSDs with built-in transparent compression to offload heavy-cost data compression operations from the CPU to the storage device, which alleviates the computing burden and removes potential interference. To the best of our knowledge, this is the first work introducing hardware-assisted data compression into a hybrid DRAM/SSD key-value cache system. 


We have implemented a prototype of ZipCache, which is a hybrid key-value cache with two cache layers, a DRAM layer and a flash memory layer. We use ScaleFlux's CSD3000 SSD~\cite{ScaleFlux-link} with hardware acceleration for transparent on-device compression. 
Our experiments show promising results. 
In comparison to the state-of-the-art solutions, including CacheLib~\cite{berg2020cachelib} and xcache~\cite{xcache-link} and Kangaroo~\cite{mcallister2021kangaroo}, 
our evaluation results demonstrate that, with a design carefully optimized for data compression, 
ZipCache can achieve up to 72.4\% higher throughput and 42.4\% lower 90-percentile read latency, and 
reduces the SSD write amplification by up to 26.2$\times$.
It is our hope that this work will motivate more future research to explore the full potential of the {\it long-overlooked} block compression for performance-critical caching systems.

The rest of this paper is organized as follows. Section~\ref{sec:background} introduces background. Section~\ref{sec:design} and~\ref{sec:evaluation} present our design and the experimental results. Section~\ref{sec:related} discusses the related work. The last section concludes this paper.

\section{Background and Motivation}
\label{sec:background}

\subsection{Data Compression}\label{sec:backgroundcompression}
\begin{figure}[htbp]
\centering 
\includegraphics[width=\linewidth]{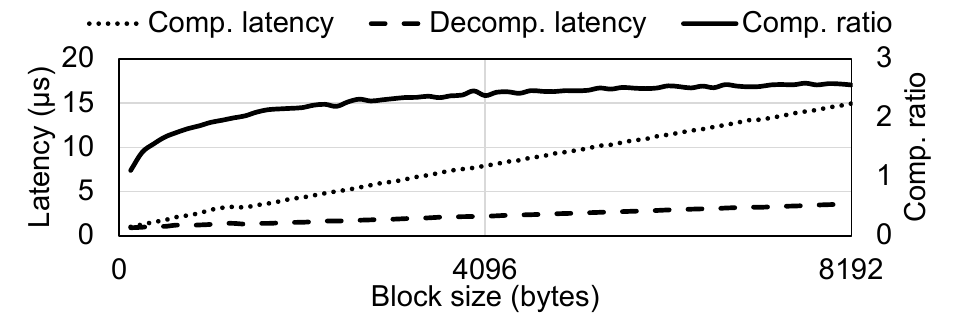}
\caption{Comparison of compression ratio, (de)compression latency under different compression block size.}
\label{fig.compress_size} 
\end{figure}

General-purpose block compression is realized by {\it deduplicating} repeated byte strings in a data block, referred to as {\it LZ search}~\cite{ziv1977universal, ziv1978compression}. Although CPU-based {\it LZ search} suffers from low speed due to the high CPU cache miss rate, its reverse process~(i.e., LZ decompression) can be much faster. 
The compression block size affects the trade-off between compression ratio\footnote{In this work, we define {\it compression ratio} as $\frac{S_{orig}}{S_{comp}}\ge 1$, where $S_{orig}$ and $S_{comp}$ denote the size of the original and compressed data blocks.} and (de)compression speed. Using the file \textit{samba} in~\textit{Silesia} corpus~\cite{silesia} as test data,
Fig.~\ref{fig.compress_size} shows the LZ4 compression ratio and (de)compression latency under different block sizes. It shows about 4$\times$ speed performance difference between compression and decompression. As the compression block size continues to increase, the compression ratio first significantly improves and then gradually saturates.

The decompression process scans through the LZ-compressed byte stream to sequentially reconstruct the original data block. In theory, this process could terminate at any byte location, leading to a {\it partially} reconstructed data block. This makes it possible to realize {\it  early termination} of decompression: Suppose {\it LZ search} compresses an $n$-segment data block ${\bf D}=[b_1, b_2,\cdots,b_n]$ into ${\bf C}$. When decompressing ${\bf C}$, the original data $[b_1, \cdots,b_k]$ are successively reconstructed as $k$ grows from 1 to $n$. Let $\tau$ denote the latency of decompressing ${\bf C}$ to reconstruct the entire ${\bf D}$. If the decompression process terminates once after the first $m$ segments $[b_1, \cdots,b_m]$, where $m\le n$)  have been reconstructed, we define its early termination factor $\gamma(m,n)=m/n$, and the decompression latency is about $\gamma(m,n)\cdot \tau$. Suppose we are only interested in obtaining one cache object in the $m$-th segment, we could reduce the latency by $1-m/n$ via decompression early termination. Using file \textit{samba} in~\textit{Silesia} corpus~\cite{silesia} as test data, we partitioned each 4KB data block into 256B~($n=16$) segments and measured the average LZ4 decompression latency under different early termination factor $\gamma(m,n)$ as shown in Table~\ref{table.partial.decompress}, which reveals substantial performance benefits. 
\begin{table}[htbp]
    \centering
    \caption{Decompression latency under different $\gamma(m,n)$.}
    \begin{tabular}{|c|c|c|c|c|c|}
    \hline
    $\gamma(m,n)$ & 1/16 & 2/16 & 4/16 & 8/16 & 16/16 \\
    \hline
    Latency ($\mu$s) &\hspace{0.1pt} 0.10 \hspace{0.1pt}& \hspace{0.1pt} 0.22 \hspace{0.1pt}&\hspace{0.1pt} 0.31 \hspace{0.1pt}&\hspace{0.1pt} 0.54 \hspace{0.1pt}&\hspace{0.1pt} 1.48 \hspace{0.1pt}\\
    \hline   
    \end{tabular}%
    \label{table.partial.decompress}
\end{table}


\subsection{Cache Index Data Structure}\label{sec:backgroundindex}
Most in-memory data stores~(e.g., Redis~\cite{Redis-link}, FASTER~\cite{chandramouli2018faster}, MICA~\cite{lim2014mica}) use hash index to reduce the latency and simplify the implementation. In contrast, most storage-based data stores~(e.g., RocksDB~\cite{RocksDB-link, dong2021rocksdb}, WiredTiger~\cite{WiredTiger-link}, Bw-tree~\cite{bwtree-13}) employ tree index to reduce the index memory usage and embrace the storage block I/O interface. As for hybrid-DRAM/SSD caches, Meta's CacheLib~\cite{berg2020cachelib, CacheLib-link} uses hash index for the DRAM and SSD tiers, while Redis-compatible xcache~\cite{xcache-link}~(developed based on Redis~\cite{Redis-link}) and Pika/RocksDB~\cite{pika-link} employs hash index for DRAM tier and log-structured merge~(LSM) tree~\cite{o1996log} index for SSD tier. 

\begin{figure}[htbp]
	\centering
 \includegraphics[width=\linewidth]{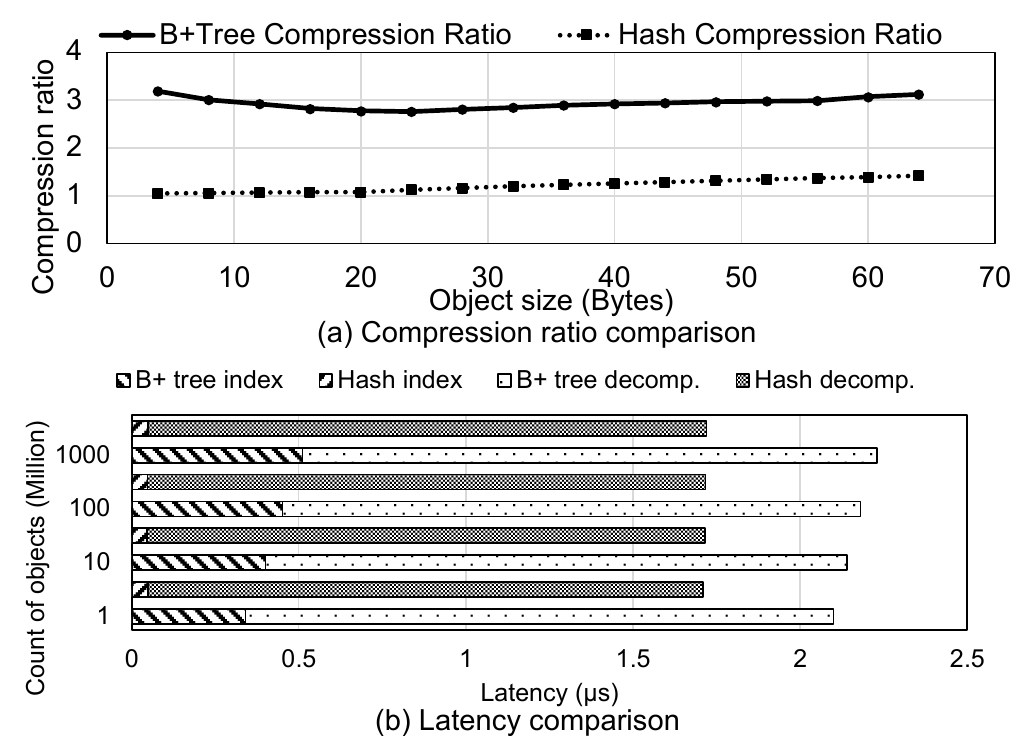}
	\caption{(a) Compression ratio of 4KB blocks under hash index and B+ tree index, and (b) latency of index traversing and block decompression under different total number of cache objects (hence different B+ tree depth), where key and values are obtained from Bitstamp Exchange Data~\cite{BTC-link}.}
	\label{fig.index_compare}
\end{figure}

The index structure has a substantial effect on the efficacy of data compression. 
Since compression ratio is proportional to the byte content similarity, a tree index that sorts all the cache objects based on their keys is clearly more beneficial, in comparison to hash index that randomly hashes cache objects into data blocks.
For the purpose of demonstration, Fig.~\ref{fig.index_compare}(a) shows the 4KB block compression ratio under hash index and B+ tree index. The key is 8B unix timestamp and object size ranging from 4B to 64B, which are both extracted from the Bitstamp Exchange Data~\cite{BTC-link}. By keeping objects with adjacent keys together, B+ tree achieves over 2$\times$ higher compression ratio than hash index. Although B+ tree index has a longer traversing latency than hash index, decompression tends to dominate the overall data access latency, which largely reduces the overall latency gap between B+ tree and hash index as shown in Fig.~\ref{fig.index_compare}(b). A strong implication is that, although prior work on in-memory cache design widely adopted hash index, integrating block compression into in-memory cache makes tree index a favorable choice. 

Meta's CacheLib uses hash index for SSD-resident objects by directly hashing each object to a 4KB SSD LBA~(logical block address) block without an in-memory hash table. 
This makes CacheLib subject to a high SSD write amplification: each object insert/update invokes re-writing a 4KB LBA block. As a variant of CacheLib, Kangaroo~\cite{mcallister2021kangaroo} applies write-ahead log~(WAL) to amortize the SSD write cost by buffering multiple cache objects hashed to the same LBA. Although it can reduce the SSD write amplification, the storage and management of WAL introduce non-negligible overhead in terms of SSD capacity and CPU/memory usage. 

\subsection{In-Storage Transparent Compression}\label{sec:introductionCSD}

\begin{figure}[htbp]
  \centering
  \includegraphics[width = 0.9\linewidth]{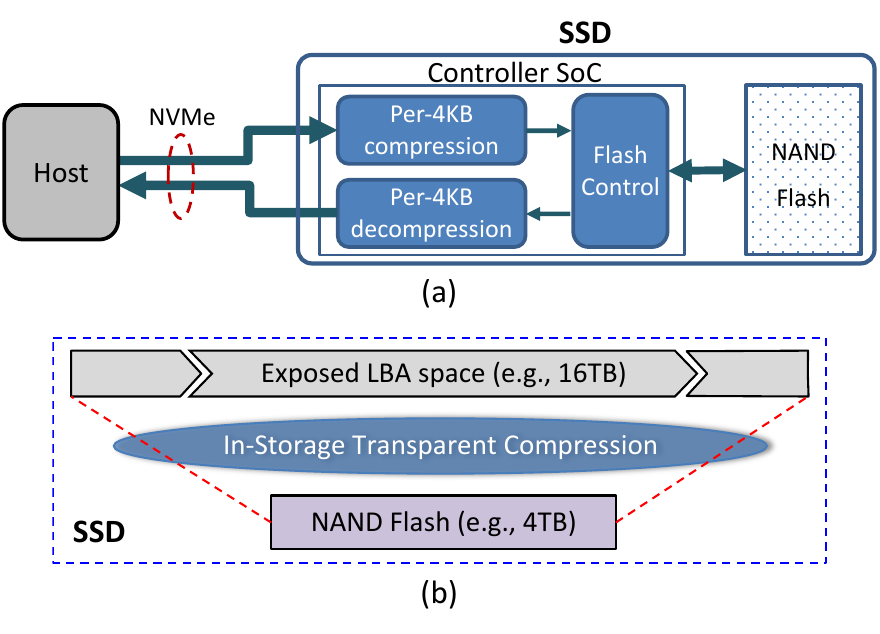}
  \caption{An illustration of (a) an SSD with built-in transparent compression, and (b) the expanded LBA space.}
  \label{fig:SSD}
\end{figure}

Data compression in a hybrid cache system imposes additional computation at both DRAM and flash cache tiers. Due to the lack of hardware support, compression over DRAM cache tier must be handled by host CPU, while the SSD-tier compression can be offloaded to the emerging computational SSDs with built-in transparent compression. Fig.~\ref{fig:SSD}(a) illustrates the structure of such SSDs~\cite{ScaleFlux-link}, where the controller SoC~(system on chip) (de)compresses each 4KB LBA block along the I/O path and manages the placement of compressed blocks on NAND flash memory. Host CPU accesses the SSD through standard I/O interface~(e.g., NVMe). The dedicated hardware engines on the controller SoC implement per-4KB (de)compression at the latency of a few microseconds, which is over 10$\times$ shorter than the TLC/QLC NAND flash memory read~(about 50$\mu$s and above) and write latency~(about 1ms and above). Therefore, SSDs with built-in transparent compression can maintain the same IOPS~(IO per second) and latency performance as ordinary SSDs. 
Such SSDs expose an expanded {\it logical} storage space that is larger~(e.g., by 2$\times$ or 4$\times$) than the physical NAND flash storage capacity, as illustrated in Fig.~\ref{fig:SSD}(b). This unique feature enables unique opportunities for our optimization efforts. 

\begin{figure*}[htbp]
\centering 
\includegraphics[width=\textwidth]{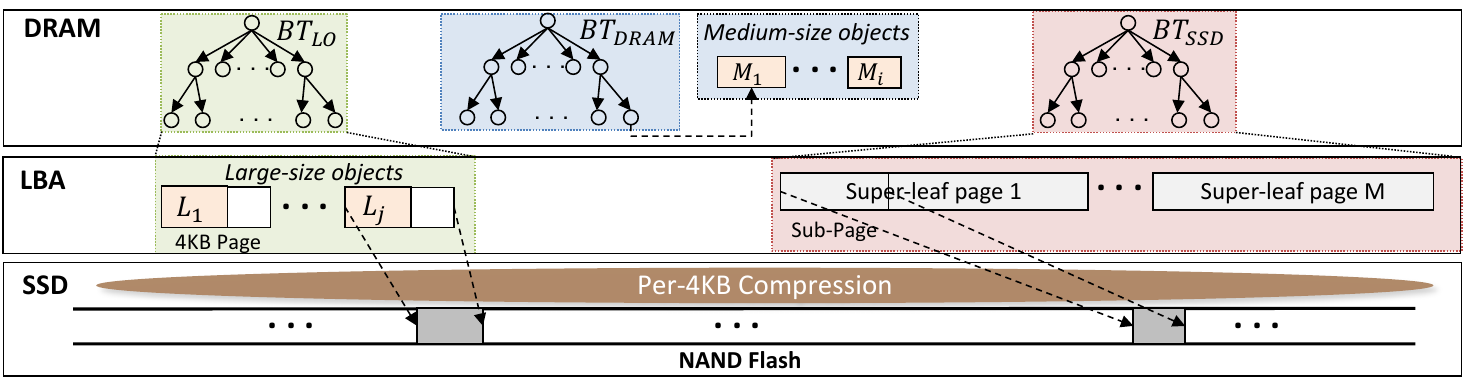} 
\caption{Overview of ZipCache architecture that employ three B+ trees to manage the DRAM cache tier, SSD cache tier, and large-size cache objects respectively.} 
\label{fig:zipcache architecture} 
\end{figure*}
\section{Design}
\label{sec:design}

We have designed a hybrid key-value cache solution, called {\it ZipCache}, with highly efficient built-in block compression. In this section, we will first introduce its basic architecture, then present a set of design techniques for improving its implementation efficiency, and finally describe its major operations.

\subsection{Architecture Overview} 
\label{subsec: indexing data structure}
ZipCache is a hybrid cache with two cache tiers. As illustrated in Fig.~\ref{fig:zipcache architecture}, ZipCache  employs B+ tree index data structure to manage both its DRAM and SSD cache tiers, performing block compression on the B+ tree leaf pages. As mentioned in Section~\ref{sec:backgroundindex}, a key benefit with B+~tree indexing is that all cache objects are {\it sorted} with their keys, enabling significantly higher compression ratios than its hash-based counterpart.

ZipCache is optimized for handling massive amount of small key-value items, which is not only practically important~\cite{berg2020cachelib} but also poses significant challenges. ZipCache categorizes cache objects into three different size classes {\it tiny}, {\it medium}, and {\it large} by using pre-defined thresholds~(e.g., 128B and 2KB). The three types of key-value items are handled differently: Tiny- and medium-size objects are stored across the DRAM and SSD tiers, while large-size objects are always SSD-resident, which is for maximizing the DRAM cache tier hit ratio. We compress in-memory tiny-size objects together in the unit of tree pages, and compress each in-memory medium-size object individually. In order to manage the different key-value items in DRAM and SSD cache tiers, ZipCache maintains three B+~trees: 
\begin{enumerate}[noitemsep,topsep=0pt]
\item $BT_{DRAM}$ for DRAM cache: This index structure entirely resides in host DRAM. Its compressed leaf pages hold tiny-size objects and pointers that point to in-memory compressed medium-size objects.
\item $BT_{SSD}$ for SSD cache: Its leaf pages hold tiny/medium-size objects and are resident in SSD, and all its non-leaf pages reside in host DRAM. 
\item $BT_{LO}$ for indexing large-size objects: It entirely resides in host DRAM, and its leaf pages hold pointers that point to SSD-resident large-size objects.
\end{enumerate}

ZipCache deploys its SSD cache tier over SSDs with built-in transparent compression, meaning that the compression of all the SSD-resident objects is transparently handled by SSDs. 
To minimize the SSD write amplification and leverage the huge DRAM-SSD bit cost gap (more than $20\times$), ZipCache adopts the inclusive caching policy over its two tiers, meaning that a key-value item could be held in both tiers. 
Due to the distinct characteristics of DRAM and SSD, the two cache tiers face different issues and challenges. Hence we will present the design of DRAM and SSD cache tiers separately in the following subsections.


\subsection{DRAM Cache Tier}
\label{subsec:DRAM}
DRAM cache tier relies on host CPUs to (de)compress each B+~tree leaf page, and a leaf page should be large enough~(e.g., 4KB) to ensure high compression ratio. 
As shown in the previous section, 
since (de)compressing a 4KB data block incurs a much higher overhead than traversing in-memory B+~tree (only a few hundred nanoseconds), we introduce the following three techniques to mitigate the (de)compression-induced overheads: 

\textbf{Decompression early termination}. 
Motivated by the observation that the decompression time is almost proportional to the amount of decompressed data, we introduce a hash-assisted method to realize early termination of the decompression process. Let ${\bf P}=[{\bf p}_1, {\bf p}_2, \cdots, {\bf p}_n]$ denote one original (uncompressed) B+ tree leaf page, and ${\bf C}$ denote the compressed version of ${\bf P}$. To obtain a cache object within the leaf page ${\bf P}$, if we {\it a priori} know that the cache object locates in the sub-page ${\bf p}_m$, we can reduce the cache read latency by roughly $1-m/n$ via decompression early termination. 

\begin{figure}[htbp]
\centering 
\includegraphics[width=\linewidth]{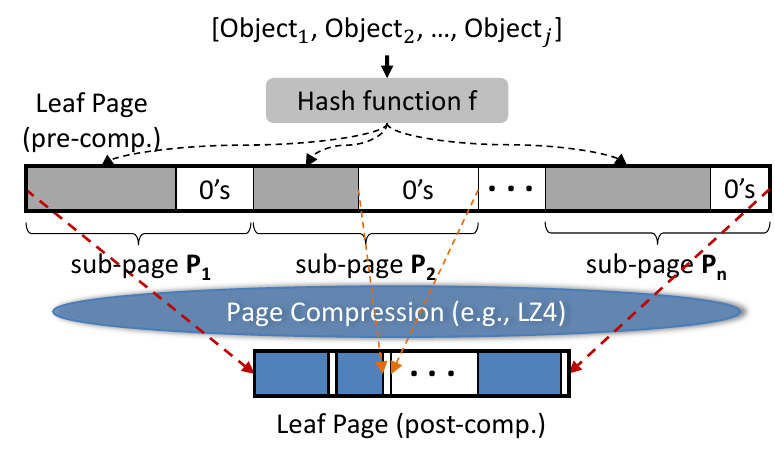}
\caption{Illustration of hash-based mapping between cache objects and sub-pages in a leaf page of $BT_{DRAM}$, including the leaf page decompression early termination.} 
\label{fig.bt.leaf}
\end{figure}

To realize decompression early termination, we must be able to know which sub-page ${\bf p}_i$ contains the requested cache object before performing decompression, which however is impossible if we construct B+~tree leaf pages with conventional practice~\cite{graefe2011modern}. To address this issue, we construct each B+ tree leaf page in a hash-based manner as illustrated in Fig.~\ref{fig.bt.leaf}: Let $\mathcal{K}$ denote the cache object key space and define a hash function $f:\ \mathcal{K}\to[1,\ n]$. For cache objects that fall into one B+ tree leaf page, we use the hash function $f$ to calculate their destined sub-pages inside the page. 
Therefore, to fetch a cache object from a B+ tree leaf page, we can determine its associated sub-page through a simple hashing and hence accordingly configure the decompression early termination. 

A side effect is that due to the nature of hashing and varying cache object size, a sub-page may only be partially filled with cache objects. However, padding the unused space with 
all zeros, our page compression process can almost entirely eliminate this potential memory space waste (see Fig.~\ref{fig.bt.leaf}). If a sub-page is completely filled up, 
the B+ tree leaf page is split into two new B+ tree leaf pages, each storing roughly half of the cache objects in the original leaf page. 

\textbf{Adaptive compression bypassing}. 
Intuitively, 
repeatedly compressing and decompressing {\it hot} B+~tree leaf pages would impose undesirable overhead, especially for workloads with highly skewed access pattern. 
We create two sub-tiers in the DRAM cache layer to adapt to runtime workloads: (1) An {\it uncompressed} sub-tier contains a small number of hot B+ tree leaf pages in their original, uncompressed form, and (2) a {\it compressed} sub-tier contains the rest leaf pages in the compressed form. To avoid sacrificing the DRAM cache tier hit ratio, we adaptively adjust the {\it uncompressed} hot-page sub-tier capacity according to the degree of workload locality. The uncompressed sub-tier could be completely eliminated in absence of sufficient locality. 

The uncompressed sub-tier contains hot data. ZipCache adaptively auto-tunes its capacity as follows. Each B+~tree leaf page is associated with a per-page counter for tracking the runtime access intensity.
The counters are periodically right-shifted by one bit to avoid overflow and age out obsolete accesses. 
Let $\mu_D$ and $\sigma_D$ denote the mean and deviation of the page access intensity, the uncompressed sub-tier only contains leaf pages whose access intensity exceeds the threshold of $\mu_D+r\cdot\sigma_D$, where $r>1$ is a design parameter. By setting $r$ sufficiently large~(e.g., 3), we can ensure only a small number of leaf pages could possibly reside in the uncompressed sub-tier. 
To further improve the adaption to runtime workloads, we may dynamically fine-tune the parameter $r$. In particular, we can vary the value of $r$ and then  monitor its effect on the overall cache performance. Optimization algorithm (e.g., simulated annealing) can be used to search the value of $r$ that better matches the runtime workload characteristics. 







\textbf{Per-page write buffering}. 
Inserting or updating a cache object in a compressed B+~tree leaf page needs to first decompress the entire page, insert/update the cache object in the page, and then re-compress the entire page. Such read/write amplification leads to a high operational overhead. We mitigate this issue by maintaining a write buffer to temporarily hold multiple insert/update to the same B+~tree leaf page and merge them together into the page through a single round of page decompression-modification-compression. This essentially trades extra memory usage for lower compression-induced implementation overhead. Once the buffer size exceeds a pre-defined threshold~(e.g., 128B or 256B), the corresponding B+~tree leaf page is marked as a candidate for background compaction. To further reduce the interference with foreground operations, the page compaction operations are handled by background threads. We note that the aggregated runtime write buffer memory usage largely depends on the spatial locality of write requests. Under a workload with high spatial locality, 
only a small percentage of leaf pages undergo intensive updates, and write buffering can effectively remove overhead caused by unnecessary re-compression.

\subsection{SSD Cache Tier} 
\label{subsec: ssd design}

The emerging computational SSD provides built-in transparent compression~\cite{ScaleFlux-link}, which brings multiple technical advantages. It not only offloads the resource-demanding (de)compression operations from the host CPUs, but also relieves the cache system from  the complexities of handling the storage of variable-length post-compression data blocks. 


Leveraging SSDs with built-in transparent compression, ZipCache SSD cache tier simply writes the B+ tree leaf pages in their original, uncompressed form into the underlying SSDs. Although being greatly assisted by such a new breed of SSDs, ZipCache SSD cache tier still needs to tackle two nontrivial issues: (1)~How to reduce its host DRAM consumption without affecting the cache hit cost? (2) How to reduce the SSD write amplification in the presence of significant size mismatch between B+ tree pages and cache objects? In the following, we present three design techniques to address these two challenging issues.

\textbf{Intra-page object hashing}. 
In order to accelerate the SSD cache hits, we keep all the non-leaf pages of the SSD-tier B+~tree in host DRAM. It allows us to access SSD only once when serving an SSD cache read request. However, this approach incurs non-trivial spatial overhead for storing these non-leaf pages in DRAM, which reduces the DRAM capacity available for DRAM cache tier. 
To mitigate this issue, the only option is to increase the size of SSD cache B+ tree leaf pages. Conventional implementation of a B+ tree always reads and writes one page as a whole on storage devices. As a result, a larger leaf page size would proportionally increase the SSD read/write amplification, leading to a higher SSD cache hit cost and shorter SSD lifetime.

\begin{figure}[htbp]
\centering 
\includegraphics[width=\linewidth]{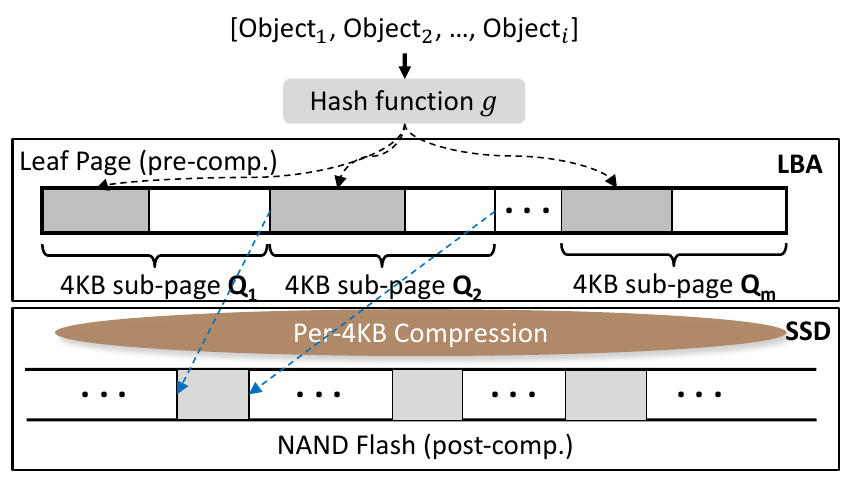}
\caption{Illustration of hash-based mapping between cache objects and 4KB sub-pages in a leaf page of $BT_{SSD}$, which decouples the the SSD read/write amplification from B+ tree leaf page size.} 
\label{fig.pt.leaf}
\end{figure}

We address the above-said challenge by decoupling the B+~tree leaf page size from SSD read/write unit. As illustrated in Fig.~\ref{fig.pt.leaf}, we construct each SSD cache B+~tree leaf page in a hash-based manner.
Since the LBA I/O size is by default 4KB, we set SSD cache's B+~tree leaf page size as a multiple of 4KB~(i.e., 4$m$~KB, where $m$ is a positive integer). Each leaf page ${\bf Q}$ is partitioned into $m$ 4KB sub-pages $[{\bf q}_1, {\bf q}_2,\cdots,{\bf q}_m]$. Let $\mathcal{K}$ denote the cache object key space and define a hash function $g:\ \mathcal{K}\to[1,\ m]$. For cache objects that fall into one B+ super-leaf page, we could use the hash function $g$ to calculate their destined 4KB sub-pages. As a result, regardless of the B+ tree page size~(e.g., 16KB or 64KB), we only need to fetch/write one 4KB from/to SSD to serve a read/write request. Such leaf pages are referred to as {\it super-leaf} pages.

Although such hash-based page construction likely leaves empty space in 4KB sub-page, we could obviate the waste of physical flash memory storage space by filling up the empty space in each 4KB sub-page with all zeros. Intra-SSD compression could seamlessly compress away the all-zero segments. It is worth noting that although both DRAM-tier and SSD-tier B+~trees adopt a similar hash-based leaf-page construction, they serve for completely different purposes: The former is for enabling decompression early termination to accelerate cache hits, while the latter is for mitigating the SSD cache read/write amplification problem by decoupling the read/write units from the B+~tree leaf pages.

\textbf{Page-based DRAM-to-SSD eviction}. 
Write amplification has a strong negative impact on SSDs in terms of both performance and lifetime. 
This and next techniques aim to reduce the write amplification caused by DRAM-to-SSD cache objects eviction. Because of intra-SSD transparent compression, we can calculate the overall write amplification as follows: 

Let $V_{obj}$ denote the total data volume of all the cache objects being evicted from DRAM to SSD, $V_{host}$ denote the total amount of data written by host to SSD through the I/O interface~(e.g., NVMe), and $V_{NAND}$ denote the total amount of data written into the NAND flash memory. We define (i) {\it host-side write amplification} $WA_{host}=V_{host}/V_{obj}\ge 1$ to represent the write amplification induced by host-side software; (ii) {\it intra-SSD write reduction} $WR_{NAND}=V_{host}/V_{NAND}\ge 1$ to quantify the effect of intra-SSD compression. Hence, we can express the overall write amplification $WA=WA_{host}/ WR_{NAND}$. To reduce the damage on NAND flash memory, we must reduce the host-side write amplification $WA_{host}$ and/or increase the intra-SSD write reduction $WR_{NAND}$.

Because both DRAM and SSD cache tiers use B+ tree index, we propose to apply page-oriented DRAM-to-SSD cache objects eviction to reduce the host-side write amplification $WA_{host}$. We choose the classical second-chance eviction policy for the purpose of implementation simplicity. Since B+~trees sort all the DRAM/SSD-resident cache objects based on their keys, the key range of a DRAM cache B+ tree leaf page may overlap with only one or few SSD cache B+~tree leaf pages. Unlike evicting small key-value items from random locations in cache, by evicting cold cache data in the unit of leaf pages, it only incurs the read-modify-write operations over a small number of SSD LBAs, leading to a small host-side write amplification $WA_{host}$. 
Moreover, thanks to the abundant spatial locality in real-world workloads, such page-based eviction 
is highly efficient, compared to object-based eviction as in CacheLib~\cite{berg2020cachelib}. 
As demonstrated in prior work~\cite{mcallister2021kangaroo} and our experiments in Section~\ref{sec:evaluation}, object-based eviction suffers from very high SSD write amplification. 

\textbf{Sub-page under-filling}. 
The objective of this technique is to reduce the SSD write reduction $WR_{NAND}$ by increasing the content compressibility of each 4KB SSD LBA block. As discussed above, within one SSD cache B+ tree super-leaf page, each cache object is hashed into one of multiple 4KB sub-pages, where each sub-page associates with one 4KB SSD LBA. Let $\beta_{fill}\le 1$ denote the 4KB sub-page fill-factor (i.e., the percentage of 4KB space that is occupied by cache objects), and the rest $1-\beta_{fill}$ portion of the sub-page is filled with all zeros. Evidently, the compression ratio of each 4KB sub-page is inversely proportional to its fill-factor $\beta_{fill}$. 
The lower the sub-page fill-factor is, the more the sub-page can be compressed inside the SSD. We set a threshold $T$ on the permissible sub-page fill-factor. As we evict pages from DRAM to SSD, 
once the fill-factor of any sub-page exceeds $T$, the entire page is split to ensure none of sub-pages have a fill-factor larger than the specified threshold. By setting $T$ well below 1~(e.g., 0.75), we can improve the sub-page compression ratio and hence increase the intra-SSD write reduction. Meanwhile, $T$ cannot be too small due to the B+ tree page split overhead. 

\subsection{Major Operations}
ZipCache supports \textsf{GET}, \textsf{SCAN}, \textsf{PUT}, and \textsf{DELETE} requests. The operation flows are summarized as follows: 

To serve a \textsf{GET} request, we search all the three B+ trees in the order of $BT_{DRAM}\to BT_{LO} \to BT_{SSD}$. Since both $BT_{DRAM}$ and $BT_{LO}$ entirely reside in host DRAM, we search the SSD cache tier B+~tree $BT_{SSD}$ in the last to ensure SSD is accessed no more than once when serving a \textsf{GET} request. If a \textsf{GET} request hits the SSD cache tier, the obtained tiny/medium-size cache object will be inserted into the DRAM cache tier. To serve a \textsf{SCAN} request, we must carry out range scans over all the three B+~trees and accordingly merge the results together as the output. 

To serve a \textsf{PUT} request, if the cache object is a tiny/medium-size object, we insert it into the DRAM cache tier, and meanwhile search the large-size object index B+ tree $BT_{LO}$ for possible cache object deletion, ensuring that any existing large-size object with the same key is removed.
If the cache object is a large-size object, we write it to SSD in the 4KB-aligned manner and insert its pointer into $BT_{LO}$, and meanwhile insert a {\it tombstone} object with the same key into the DRAM cache tier to perform possible cache object deletion. 
We note that a {\it tombstone} inserted into the DRAM cache tier will not disappear until it reaches the SSD cache tier. To serve a \textsf{DELETE} request, we insert one {\it tombstone} object into the DRAM cache tier, and search $BT_{LO}$ for possible cache object deletion. 

Besides normal B+ tree management operations such as page split, ZipCache carries out two additional major background operations: (1) Leaf page re-compression in DRAM cache B+~tree: 
As DRAM cache B+ tree uses per-page write buffering to amortize the leaf page re-compression cost, once the size of one per-page write buffer reaches the pre-specified threshold, ZipCache performs background decompression-modify-compression over the leaf page to merge the buffered objects into the compressed leaf page. (2) DRAM-to-SSD page eviction: 
ZipCache keeps track of the 
hotness of each DRAM cache B+ tree leaf page. When the DRAM cache tier runs out of memory space, ZipCache evicts cold DRAM-resident leaf pages into the SSD cache tier. All the in-memory medium-size objects associated with to-be-evicted pages are first decompressed and then moved to the SSD cache tier together with other tiny-size objects.

\section{Evaluation}
\label{sec:evaluation}
We have implemented the ZipCache prototype in C++ and carried out experiments on a server with two Intel Xeon Gold 6134 CPUs, 384GB DRAM, and one 3.84TB ScaleFlux CSD 3000 drive with built-in transparent compression. The system OS is Ubuntu Linux release 22.04. Being fully compliant with the NVMe protocol, the CSD 3000 drive realizes hardware-based zlib (de)compression on each 4KB LBA data block along the I/O path. It can
achieve a compression ratio similar to that of the software zlib library at the level 6, and its (de)compression latency is sub-5$\mu s$, at least one order of magnitude shorter than TLC/QLC NAND flash memory read/write latency. 
Meta's Cachebench~\cite{berg2020cachelib} is used to generate realistic cache object access workloads with the configurable access locality. By configuring the percentage of cache objects that collectively serve 80\% cache access requests in Cachebench, we carried out experiments based on four different categories of workload locality as listed in Table~\ref{table.locality}, where 80\%$\to$20\% means that 80\% cache access requests hit 20\% of all the cache objects. In the case of {\it zero locality}, we configure Cachebench to randomly generate requests over all the cache objects, representing the worst-case scenario. 
\begin{table}[htbp]
    \centering
    \caption{Four categories of workload locality.
    }
    \begin{tabular}{|c|c|c|c|}
    \hline
    Strong & Moderate & Weak & Zero\\
    \hline   
    80\%$\to$8\% & 80\%$\to$20\% & 80\%$\to$64\% & \textit{Random}\\
    \hline
    \end{tabular}%
    \label{table.locality}
\end{table}

To study the impact of cache object content compressibility, we modified Cachebench to generate the cache object content as follows. Given a parameter $\eta\in[0\%,100\%)$, we fill $1-\eta$ of each cache object with incompressible random content and set the remaining $\eta$ portion as all zeros. Hence, the cache object content compressibility improves as $\eta$ increases. The same method has been used by the popular I/O test tool FIO~(flexible I/O tester)~\cite{FIO-link} to generate I/O data with configurable compressibility.


\subsection{Overall Cache Performance}

\begin{figure*}[htbp]
	\centering
\includegraphics[width=\linewidth]{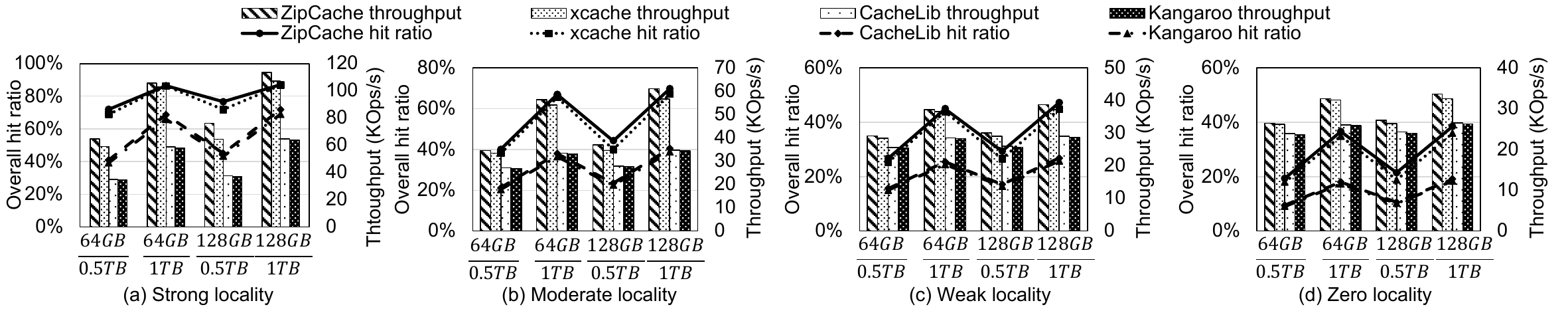}
\caption{Comparison of throughput and overall DRAM/SSD cache hit ratio when the active working set is much larger than the DRAM/SSD cache capacity. We fixed the total active working set size $C_{WS}$ as 6TB and, under each category of workload locality, considered four different settings of \{$C_{DRAM}$, $C_{SSD}$\}: \{64GB, 0.5TB\}, \{64GB, 1TB\}, \{128GB, 0.5TB\}, and \{128GB, 1TB\}.}
	\label{fig.overall experiment}
\end{figure*}

\begin{figure*}[htbp]
	\centering
\includegraphics[width=\linewidth]{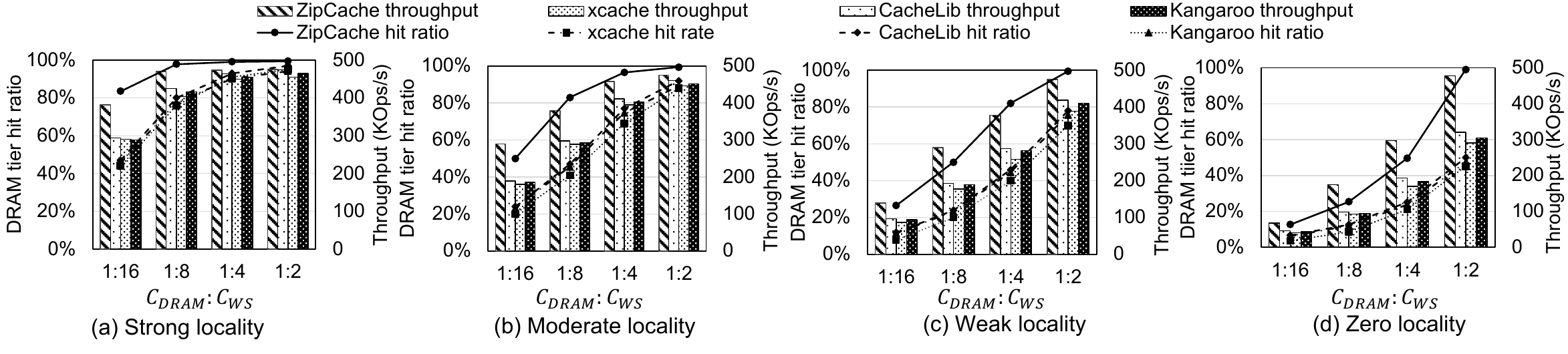}
\caption{Comparison of throughput and DRAM cache hit ratio when the active working set fits into the DRAM/SSD cache~(hence the SSD cache tier hit ratio is 100\%). We fixed the DRAM cache tier capacity $C_{DRAM}$ as 64GB and considered $C_{DRAM}:C_{WS}$ ratio of 1:16, 1:8, 1:4, and 1:2.}
	\label{fig.throughput_hitratio}
\end{figure*}

We first evaluate and compare the  speed performance of CacheLib~\cite{CacheLib-link}, Kangaroo~\cite{mcallister2021kangaroo}~(a variant of CacheLib for reducing SSD write amplification), xcache~\cite{xcache-link}, and ZipCache. 
To cover a wide range of spectrum, we considered the scenarios when the total active working set size is either larger or smaller than the DRAM/SSD cache capacity. For both scenarios, we set the key size as 16B and cache object size as 64B. The cache object content is generated with the compressibility parameter $\eta=50\%$. The workload consists of 16 user threads issuing \textsf{GET} and \textsf{PUT} requests at the ratio of 1:1. Since adaptive compression bypassing is effective only under workloads with very strong locality, we disable this feature here an will study its effect later in Section~\ref{sec:adaptive}. The parameters of CacheLib and xcache are set as their default values, and we turned on the LZ4 compression of xcache's SSD cache tier. For ZipCache, the leaf page size of DRAM tier B+~tree and SSD tier B+~tree is set to 4KB and 64KB, respectively, and the DRAM tier per-page write buffer size limit is set to 256B.

Fig.~\ref{fig.overall experiment} shows the throughput and overall cache hit ratio when the active working set is much larger than the DRAM/SSD cache capacity. We fixed the total active working set size $C_{WS}$ as 6TB. Let $C_{DRAM}$ and $C_{SSD}$ denote the DRAM and SSD cache tier capacity, we considered four different settings of \{$C_{DRAM}$, $C_{SSD}$\}: \{64GB, 0.5TB\}, \{64GB, 1TB\}, \{128GB, 0.5TB\}, and \{128GB, 1TB\}. In case of the SSD cache tier miss due to the larger-than-cache active working set, we configured the backend data access latency as 1ms. 
Since xcache/ZipCache both apply compression over the SSD cache tier, they have similar overall cache hit ratio that is higher than CacheLib. As a result, xcache and ZipCache have higher throughput than CacheLib and Kangaroo, as shown in Fig.~\ref{fig.overall experiment}. Meanwhile, since ZipCache applies compression over its DRAM cache tier, it has a higher DRAM cache tier hit ratio (hence higher throughput) than xcache. For instance, under moderate workload locality with \{$C_{DRAM}$, $C_{SSD}$\} of \{128GB, 1TB\}, the overall cache hit ratio of ZipCache is 69.7\%, 2.6 percentage points (p.p.), 29.2 p.p. and 30.9 p.p. higher than xcache, CacheLib, and Kangaroo.
The throughput of ZipCache is about 6.1\%, 75.0\% and 76.1\% than that of xcache, CacheLib and Kangaroo, respectively. As shown in Fig.~\ref{fig.overall experiment}, as the workload locality weakens, the performance difference among the three caches becomes smaller since the backend access latency becomes more dominant.




Fig.~\ref{fig.throughput_hitratio} shows the throughput and DRAM cache tier hit ratio when the active working set fits completely in the DRAM/SSD cache~(hence the SSD cache tier hit ratio is 100\%). We fixed the DRAM cache tier capacity $C_{DRAM}$ as 64GB and considered the $C_{DRAM}:C_{WS}$ ratio of 1:16, 1:8, 1:4, and 1:2. Because of the build-in block compression over its DRAM cache tier, ZipCache consistently achieves higher DRAM tier hit rate and higher throughput than CacheLib, Kangaroo and xcache, with throughput improvements of up to 72.4\%. Compared with xcache, CacheLib achieves slightly higher DRAM cache hit ratio and hence throughput than xcache. 
Under high workload locality~(e.g., strong/moderate locality), the advantage of ZipCahe over CacheLib/Kangaroo/xcache gradually diminish as the DRAM capacity increases. For example, 
under workloads with {\it strong locality}, when a small cache capacity (${C_{DRAM}}:{C_{WS}}$ of 1:16), ZipCache achieves about 36.3 p.p. higher DRAM tier hit rate 
and 30.4\% higher throughput than CacheLib/Kangaroo/xcache; when ${C_{DRAM}}:{C_{WS}}$ increases to 1:2, their DRAM hit rate and throughput become almost the same. This is because, under high workload locality, increasing the DRAM capacity alone~(without data compression) can be sufficient to quickly raise the DRAM tier hit rate over 90\%. In comparison, under relatively low workload locality~(e.g., weak/zero locality), increasing the DRAM capacity alone is much less effective on improving the DRAM tier hit ratio. As a result, the value of data compression becomes more evident. For example, 
with {\it zero locality}, ZipCache achieves a DRAM hit rate that is 6.4 to 49.2 p.p. higher than CacheLib, when ${C_{DRAM}}:{C_{WS}}$ increases from 1:16 to 1:2. 


\subsection{DRAM Cache Tier Compression}

\begin{figure}[htbp]
	\centering
    \includegraphics[width=\linewidth]{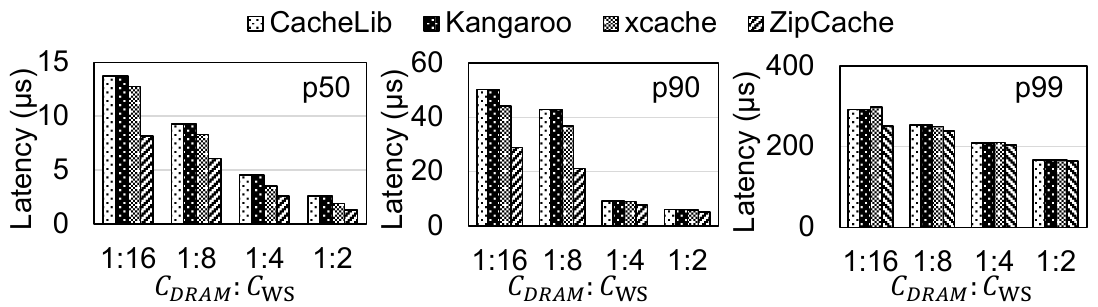}
	\caption{\textsf{GET} latency under workloads with moderate locality and different $C_{DRAM}:C_{WS}$.}
\label{fig.latency_percentile}
\end{figure}

Fig.~\ref{fig.latency_percentile} further shows the \textsf{GET} latency when active working set fits in DRAM/SSD cache, where we measured the 50-percentile~(p50) latency, 90-percentile~(p90 latency), and 99-percentile~(p99) latency. Compared to CacheLib, Kangaroo, and xcache, ZipCache consistently achieves shorter latency due to its higher DRAM cache tier hit ratio, with latency reductions of up to 42.4\%.
Moreover, the latency reduces as the DRAM cache tier capacity increases, e.g., the p90 latency of ZipCache reduces from 28.9$\mu$s to 5.2$\mu$s when ${C_{DRAM}}:{C_{WS}}$ increases from 1:16 to 1:2. All the three have similar p99 latency since it is  largely determined by the SSD tier read latency.

\begin{figure}[htbp]
	\centering
    \includegraphics[width=\linewidth]{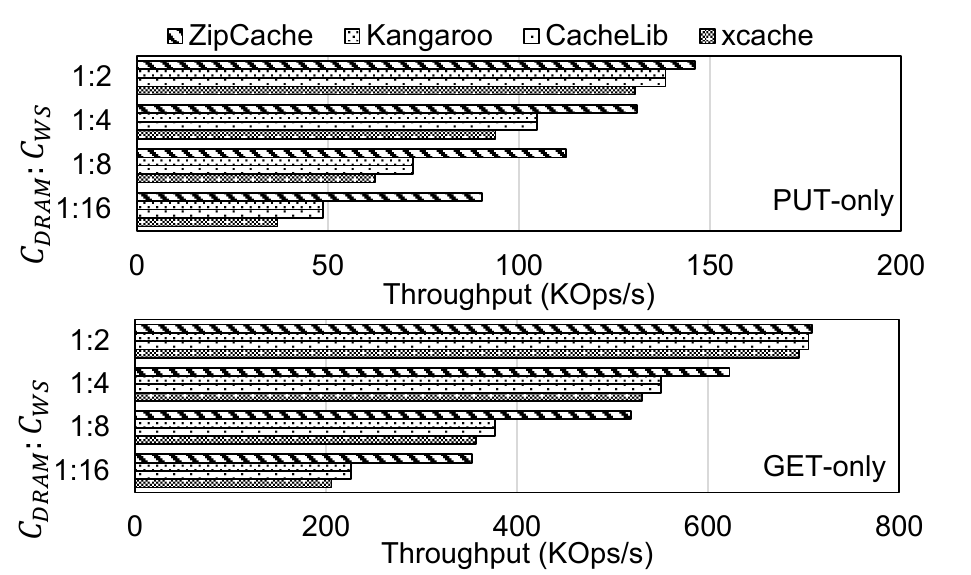}
	\caption{Average throughput of \textsf{PUT}-only and \textsf{GET}-only workload with moderate locality and different $C_{DRAM}:C_{WS}$.}
\label{fig.throughput of get and put only}
\end{figure}

In addition to workloads with \textsf{GET}:\textsf{PUT} ratio of 1:1, we further compared the performance under \textsf{PUT}-only and \textsf{GET}-only Cachebench workloads with moderate locality.
As shown in Fig.~\ref{fig.throughput of get and put only}, under $C_{DRAM}:C_{WS}$ of 1:16 and \textsf{PUT}-only workload, ZipCache achieves 78.1\% and 145.7\% higher throughput than CacheLib/Kangaroo and xcache, because its DRAM tier compression could help reducing the SSD tier write amplification due to DRAM-to-SSD eviction. Under $C_{DRAM}:C_{WS}$ of 1:16 and \textsf{GET}-only workload, ZipCache achieves 56.3\% and 71.6\% higher throughput than CacheLib/Kangaroo and xcache, because its DRAM tier compression helps to increase the DRAM tier hit ratio. The performance difference gradually shrinks as the DRAM cache tier capacity increases. 

The above results well demonstrate the effectiveness of incorporating in-memory data compression to improve the cache speed performance. This essentially attributes to the substantially increased DRAM cache tier hit ratio enabled by in-memory compression and the significant data access latency gap between DRAM and SSD.  

\subsection{SSD Cache Tier Write Amplification}
\label{sec:eval:ssd}
We further compared the SSD cache tier write amplification among ZipCache, CacheLib~\cite{CacheLib-link}, Kangaroo~\cite{mcallister2021kangaroo}, and xcache~\cite{xcache-link}. As discussed above in Section~\ref{subsec: ssd design}, once we deploy a hybrid-DRAM/SSD cache on SSDs with built-in compression, we can express the overall SSD write amplification as $WA=WA_{host}/ WR_{NAND}$, where the {\it host-side write amplification} $WA_{host}\ge 1$ represents the write amplification induced by host-side cache software and  {\it intra-SSD write reduction} $WR_{NAND}\ge 1$ quantifies the effect of intra-SSD compression~(i.e., compression ratio achieved by SSD). 
\begin{figure}[t]
	\centering
	\subfigure[Host-side write amplification] {\includegraphics[width=\linewidth]{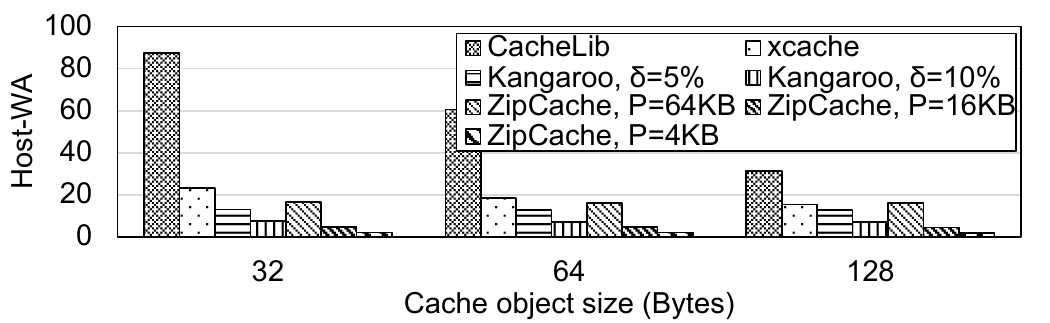}
  \label{fig.waf_host}}
	\subfigure[Intra-SSD write reduction] {\includegraphics[width=\linewidth]{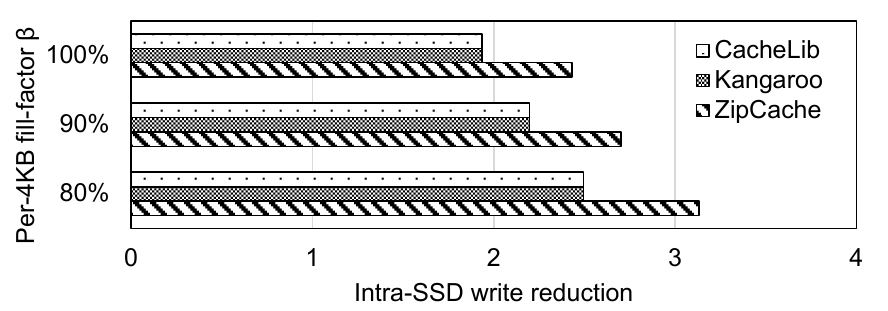}
 \label{fig.waf_nand}}
	\caption{(a) Host-side write amplification $WA_{host}$ comparison among CacheLib, Kangaroo and ZipCache under different cache object size, and (b) intra-SSD write reduction under different per-4KB fill-factor~(xcache is not included since its intra-SSD write reduction remains as 1).}
	\label{fig.waf}
\end{figure}

Fig.~\ref{fig.waf_host} shows the host-side write amplification $WA_{host}$ under the Cachebench workload with {\it moderate locality}. The key size is 16B and cache object size is 32B, 64B, and 128B.  By directly hashing each cache object into one SSD 4KB LBA, CacheLib experiences very high $WA_{host}$ that is inversely proportional to the cache object size. For example, its $WA_{host}$ increases from 31.5 to 60.7 when the cache object size reduces from 128B to 64B. Regarding xcache, its DRAM cache uses hash index and SSD cache uses LSM-tree index. Hence, xcache has lower SSD write amplification than CacheLib. By complementing CacheLib with a WAL to accumulate multiple cache objects hashed to the same SSD 4KB LBA,   Kangaroo~\cite{mcallister2021kangaroo} reduces $WA_{host}$ at the cost of SSD storage capacity. Let $\delta<1$ denote the ratio of WAL size and SSD cache size, we considered two values of $\delta$: 5\% and 10\%. Fig.~\ref{fig.waf_host} shows the effectiveness of WAL-assisted write amplification and the trade-off between the WAL-induced storage overhead $\delta$ and host-side write amplification $WA_{host}$. For ZipCache, we considered three leaf page size of SSD tier B+ tree $BT_{SSD}$, including 4KB, 16KB, and 64KB. As discussed above, to minimize SSD tier cache hit time, $BT_{SSD}$ keeps all its non-leaf pages in DRAM and leaves leaf pages on SSD. We define the $BT_{SSD}$ memory overhead (denoted as $\zeta$) as the ratio between the total size of its in-memory non-leaf pages and total size of its on-SSD leaf pages. The $BT_{SSD}$ leaf page size affects the trade-off between host-side write amplification $WA_{host}$ and $BT_{SSD}$ memory overhead $\zeta$, which can be observed from Fig.~\ref{fig.waf_host} and  Table~\ref{table.waf}. The results show that, dependent upon their different configurations, Kangaroo and ZipCache have comparable host-side write amplification, which is slightly better than xcache and significantly better than that of CacheLib, especially under small cache object size. 
\begin{table}[htbp]
    \centering
    \caption{$BT_{SSD}$ memory overhead $\zeta$.}
    \resizebox{0.4\textwidth}{!}{%
    \begin{tabular}{|c|c|c|c|c|}
    \hline
        Leaf page size &\hspace{3pt} 64KB \hspace{3pt}&\hspace{3pt} 16KB \hspace{3pt}& \hspace{3pt} 4KB \hspace{3pt}\\ \hline
        \hspace{3pt} Memory overhead \hspace{3pt}$\zeta$ & 0.6\% & 2.1\% & 7.5\% \\ \hline
    \end{tabular}
    }
    \label{table.waf}
\end{table}

Fig.~\ref{fig.waf_nand} shows the measured intra-SSD write reduction $WR_{NAND}$ when running CacheLib, Kangaroo, and ZipCache on SSD with built-in transparent compression. Since xcache's LSM-tree-based SSD cache tier applies block compression, it does not benefit from intra-SSD transparent compression and hence always has the intra-SSD write reduction of 1. We generate the content of each cache object with the compressibility parameter $\eta=50\%$. As discussed above, the per-4KB under-filling technique can be equally applied to CacheLib and Kangaroo. Hence, we measured the intra-SSD write reduction of all three caches under different per-4KB fill-factor $\beta_{fill}$ including 100\%, 90\%, and 80\%. Since Kangaroo and CacheLib employ the same hash-based SSD tier cache structure, they have the same intra-SSD write reduction $WR_{NAND}$. As discussed above in Section~\ref{sec:backgroundindex}, by sorting all the cache objects based on their keys, B+ tree leaf pages have a higher compressibility due to the stronger content correlation across adjacent keys. Hence, ZipCache can achieve larger intra-SSD write reduction. For example, under the fill-factor of 80\%, CacheLib/Kangaroo have the $WR_{NAND}$ of 2.49 while ZipCache has $WR_{NAND}$ of 3.13. 

By combining the above results of host-side write amplification and intra-SSD write reduction, we can observe that, in addition to its higher DRAM tier hit ratio and hence higher speed performance, ZipCache achieves significantly lower SSD write amplification compared to CacheLib and xcache, with reductions of up to 26.2$\times$. For example, with the object size of 64B and per-4KB fill-factor of 90\%, the overall SSD write amplification of ZipCache~(leaf page size of 16KB) is 1.8, which is only 3.7\% and 9.8\% of CacheLib and xcache, respectively. Even compared with Kangaroo, the CacheLib variant that is solely optimized for reducing the SSD write amplification, ZipCache has comparable or lower SSD write amplification. For example, with the object size of 64B and per-4KB fill-factor of 90\%, the overall SSD write amplification of ZipCache~(leaf page size of 16KB) is 53.5\% of Kangaroo~(with storage space overhead $\delta$ of 10\%).   


\subsection{Adaptive Compression Bypassing}\label{sec:adaptive}
For workloads with strong localities, we could noticeably improve the cache performance by adaptively bypassing the compression over very hot leaf pages of the DRAM-tier cache's B+~tree $BT_{DRAM}$. In this section we will study the effect of such adaptive compression bypassing. 
\begin{figure}[htbp]
	\centering
\includegraphics[width=\linewidth]{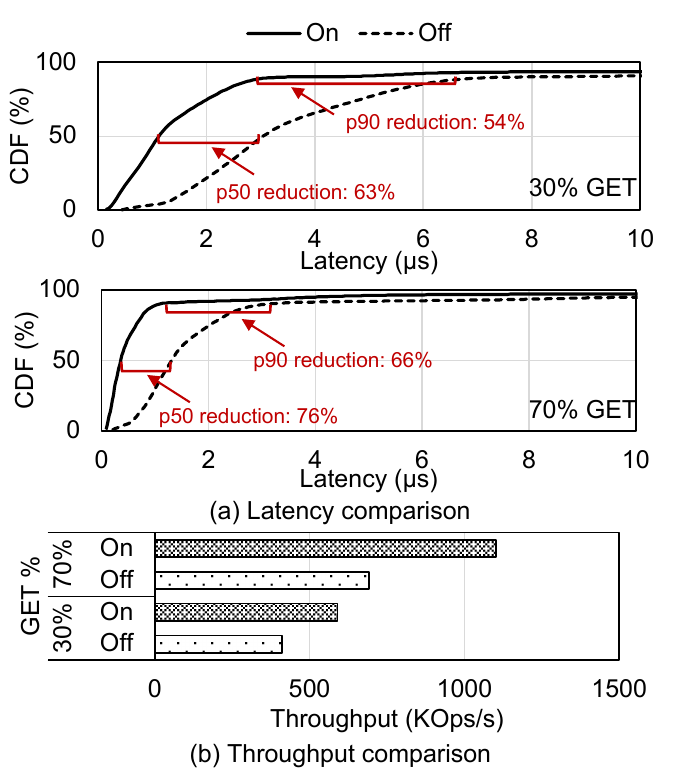}
	\caption{(a) \textsf{GET} latency and (b) average throughput comparison under Cachebench workload with {\it strong locality}, where \textsf{GET}:\textsf{PUT} ratio is 30\%:70\% or 70\%:30\%.}
	\label{fig.adaptive compression bypassin}
\end{figure}
Fig.~\ref{fig.adaptive compression bypassin} shows the CDF~(cumulative distribution function) of the measured \textsf{GET} latency and average throughput under the Cachebench workload with {\it strong locality}. 
We configure a write-intensive and a read-intensive workload with the \textsf{GET}:\textsf{PUT} ratio set to 30\%:70\% and 70\%:30\%, respectively. The results show that the proposed \textit{adaptive compression bypassing} can significantly improve the throughput and reduce the perceived latency. For example, by turning on \textit{adaptive compression bypassing}, we could reduce the p50 \textsf{GET} latency by 63\% for the write-intensive workload
and 76\% for the read-intensive workload,
correspondingly improving the average throughput by 44\% and 59\%, respectively.
The results clearly show the benefits of obviating CPU-intensive (de)compression operations over hot B+~tree leaf pages in the DRAM cache tier. 

\begin{figure}[htbp]
\centering 
\includegraphics[width=\linewidth]{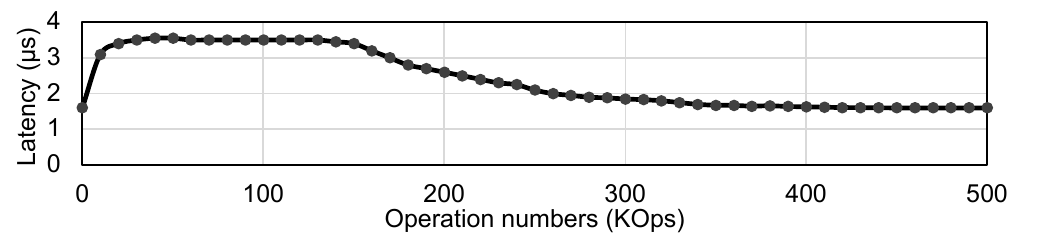} 
\caption{Change of the average \textsf{GET} latency after the sudden \textit{hot region} position shift.} 
\label{fig.position} 
\end{figure}

In above experiments, the workload {\it hot region} remains stationary and hence has been fully captured by ZipCache. We further carry out experiments to study the responsiveness of compression bypassing to runtime workload variations. Using the same Cachebench workload with {\it strong locality}, we arbitrarily shift the position of the {\it hot region} to cover a non-overlapping set of in-memory cache objects. Fig.~\ref{fig.position} shows the measured average \textsf{GET} latency before and after this sudden  hot region shift. In the figure, we can see that the average \textsf{GET} latency rapidly increases from 1.6$\mu$s to 3.4$\mu$s due to the working-set change. As compression bypassing responds to the workload change by re-compressing the pages of the previous working set and decompressing the pages of the new hot region, the average latency gradually returns back to 1.6$\mu$s after serving about 500K operations. Given the average throughput of over 400K operations per second under strong locality (as shown above in Fig.~\ref{fig.throughput_hitratio}), we can estimate the transition period of no more than few seconds. 

\subsection{Sensitivity Study}

\begin{figure}[htbp]
	\centering
        \includegraphics[width=\linewidth]{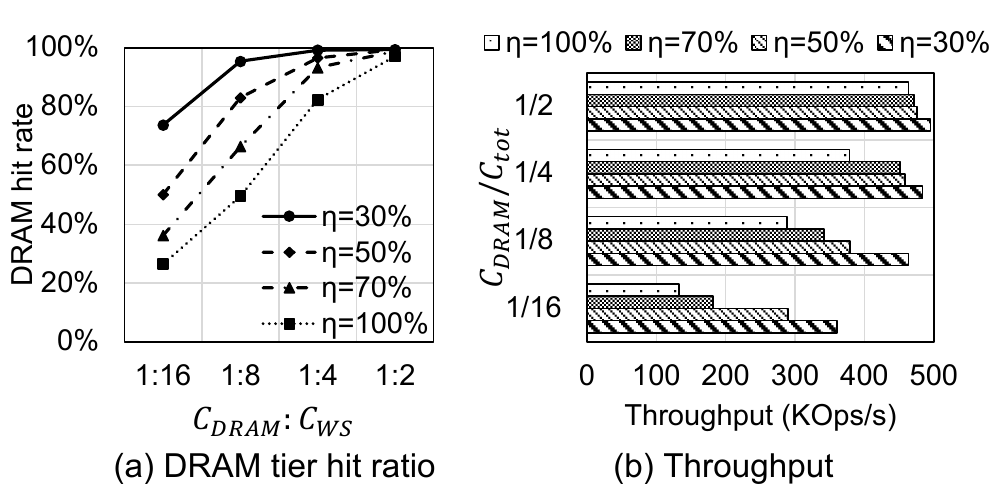}
	\caption{(a) DRAM tier cache hit ratio and (b) overall throughput under different compressibility parameter $\eta$.}
	\label{fig.hitratio_compressionratio}
\end{figure}

{\bf Compressibility}. ZipCache has several configurable parameters. 
In the above experiments, we fixed the cache object compressibility parameter $\eta$ as 50\%. Fig.~\ref{fig.hitratio_compressionratio} shows the measured DRAM cache tier hit ratio and throughput under different settings of $\eta$. We use the Cachebench workload with {\it moderate locality} and set the key size to 16B and cache object size to 64B. The results show the effect of data content's compressibility on the cache performance. With the DRAM cache capacity vs.~active working set size ratio $C_{DRAM}:C_{WS}$ of 1:16, as the compressibility increases from incompressible~($\eta=100\%$) to highly compressible ($\eta=30\%$), the DRAM tier cache hit ratio increases from 26.4\% to 73.6\%. The results clearly show the impact of data compressibility on the performance of ZipCache.

\begin{figure}[htbp]
	\centering
        \includegraphics[width=\linewidth]{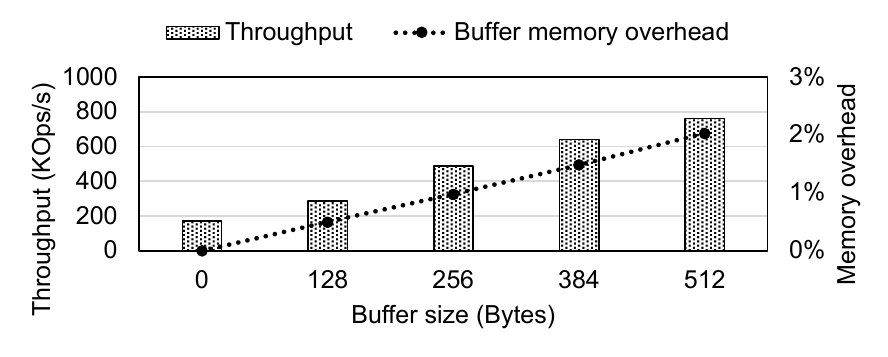}
	\caption{Throughput and memory usage overhead under different per-page write buffer size, where \textsf{PUT}-only Cachebench workload with \textit{strong locality} is used.}
	\label{fig.throughput_buffer_size}
\end{figure}

{\bf Write Buffer Size}. The per-page write buffer size could also noticeably impact the DRAM tier cache performance under write-intensive workloads. As we reduce the per-page write buffer size to save DRAM space usage, the DRAM-tier cache's B+~tree would experience more frequent page re-compression, leading to degraded performance. Fig.~\ref{fig.throughput_buffer_size} shows the ZipCache throughput and DRAM usage overhead under different per-page write buffer sizes using the \textsf{PUT}-only Cachebench workload with {\it strong locality}. The overhead of DRAM usage is defined as the ratio between the aggregated per-page write buffer size and total DRAM tier cache capacity. The results clearly show the trade-off between the cache  performance and the overhead of DRAM usage. As we increase the per-page write buffer size from 0 to 256B, the cache performance improves by 2.9$\times$ at the cost of 1\% DRAM usage overhead.

\begin{figure}[htbp]
	\centering
        \includegraphics[width=\linewidth]{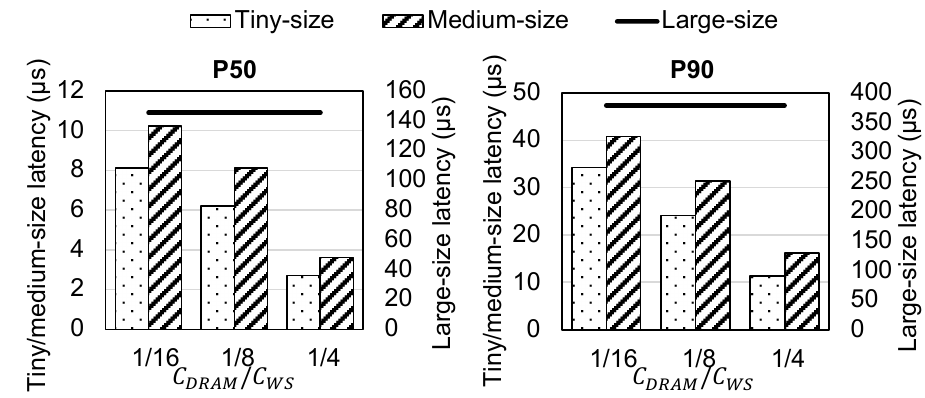}
	\caption{\textsf{GET} latency for various sizes objects under the Cachebench workload with \textit{modest locality}.} 
	\label{fig.hybrid_size_get_latency}
\end{figure}

{\bf Cache Object Size}. Given the same total cache capacity, as cache object size decreases, the number of cache objects increases, leading to more cache index implementation complexity and more significant compression-induced overhead. Hence, to most heavily stress the cache, the experiments above focus on workloads with only tiny-size cache objects. To show the effect of cache object size, we have performed experiments with Cachebench workloads that contain tiny-size~(64B), medium-size~(256B), and large-size~(2KB) cache objects. We ensure that the three categories of cache objects consume the same amount of cache capacity. Fig.~\ref{fig.hybrid_size_get_latency} shows the \textsf{GET} latency for the three different sizes of cache objects. Requests over medium-size objects experienced longer latency than that over tiny-sized objects. For instance, under $C_{DRAM}:C_{WS}$ of 1:16, the p50 latency for medium-size objects is 26\% more than that for tiny objects. This increased latency is due to the extra step of decompressing these medium-size objects. Large-size objects, stored on SSDs, have significantly higher \textsf{GET} latency compared to both tiny and medium-size objects.

\section{Related Work}
\label{sec:related}
\noindent\textbf{Key-Value Stores}: The design and implementation of in-memory and SSD-based key-value stores have been extensively studied by the research community. Most in-memory KV stores chose to employ hash-based index~(e.g., Memcached~\cite{Memcached-link}, Redis~\cite{Redis-link}, MemC3~\cite{fan2013memc3}, MICA~\cite{lim2014mica}, Mega-KV~\cite{zhang2015mega}, RAMCloud~\cite{ousterhout2015ramcloud}, FASTER~\cite{chandramouli2018faster}, HotRing~\cite{chen2020hotring}). Only few implementations, such as MassTree~\cite{mao2012cache}, use tree-based index. 
SSD-based key-value stores typically employ tree-based index and support block data compression. Because of its low write amplification, log-structured merge~(LSM) tree~\cite{o1996log} index data structure has received most recent attentions in research community for building SSD-based key-value stores~\cite{RocksDB-link, PebblesDB-17, balmau2017triad, dayan2018dostoevsky, huang2019x, chen2021spandb}. 

Prior work also studied the design of hybrid key-value stores over different memory technologies~(e.g., DRAM and SSD), aiming to strike a better balance between the speed and cost. CacheLib~\cite{berg2020cachelib} and xcache~\cite{xcache-link} are two representative hybrid-DRAM/SSD key-value stores, where CacheLib uses hash-based index for both the DRAM and SSD tiers and xcache uses hash-based index for DRAM tier and LSM-tree index for SSD tier. Motivated by intensive research on NVM~(non-volatile memory) technologies over the past decade, recent work has developed NVM-enhanced hybrid key-value stores~\cite{zhan2020rangekv, yao2020matrixkv, ding2022trianglekv, duan2023revisiting}. Though there is prior research on flash cache compression~\cite{li2016cachededup,wang2020austere,li2014nitro,hao2021implementing}, little work studied the potential of improving cache performance via in memory block data compression.

\noindent\textbf{Memory Compression}: The computer architecture research community has widely studied the implementation of hardware-based main memory compression~\cite{choukse2018compresso,ekman2005robust,pekhimenko2013linearly,zhao2015buri}). Aiming at better serving general-purpose computing systems, hardware-based main memory compression focuses on fine-grained per-cacheline~(e.g., 64B) compression. The Linux kernel feature Zswap~\cite{Zswap-link} compresses to-be-swapped 4KB pages and keeps them in DRAM, which has been used by Google~\cite{lagar2019software} and Meta~\cite{weiner2022tmo} to increase effective DRAM capacity. Contemporary data analytics systems like SAP HANA~\cite{farber2012sap}, Oracle~\cite{lahiri2015oracle}, and Snowflake~\cite{dageville2016snowflake} apply block compression to their in-memory column-stores to reduce their memory consumption. In-memory time series databases~\cite{adams2020monarch,pelkonen2015gorilla} also widely use compression to exploit the inherently high compressibility of time series data.

\noindent\textbf{Computational SSD}: Computational storage has attracted significant recent interest~\cite{cao2020polardb, barbalacecomputational-21, kwon2021fast, vinccon2022near}, and commercial products are emerging on the commercial market~(e.g., Samsung's SmartSSD~\cite{SmartSSD-link} and ScaleFlux's CSD~\cite{ScaleFlux-link}). Recent research~\cite{Zheng-Hotstorage-20, chen2021kallaxdb, huang2023breathing} has studied how database management systems could take advantage of computational SSDs with built-in transparent compression. 

\section{Conclusion}
This paper presents a hybrid cache design called ZipCache, which integrates block data compression to improve the cache performance. To maximize the block compression ratio and hence cache hit ratio, in contrast to most existing in-memory cache design, ZipCache employs the classic B+ tree indexes to manage both the DRAM and SSD cache tiers. We developed several design techniques to reduce compression-induced DRAM tier cache hit cost overhead. Built upon emerging SSDs with transparent compression, ZipCache SSD tier cache incorporates several design techniques to reduce the its B+~tree index memory consumption and reduce the SSD write amplification, especially for workloads dominated by tiny-size cache objects. Extensive experiments demonstrated its effectiveness and studied the involved design trade-offs.

\begin{acks}

This work was supported in part by the National Science Foundation (NSF) under grants CNS-2006617, CCF-1910958, CCF-2210754, CCF-2210755, CCF-2312508, and CCF-2312509.
\end{acks}
\bibliographystyle{ACM-Reference-Format}
\bibliography{reference}


\begin{thebibliography}{65}


\ifx \showCODEN    \undefined \def \showCODEN     #1{\unskip}     \fi
\ifx \showDOI      \undefined \def \showDOI       #1{#1}\fi
\ifx \showISBNx    \undefined \def \showISBNx     #1{\unskip}     \fi
\ifx \showISBNxiii \undefined \def \showISBNxiii  #1{\unskip}     \fi
\ifx \showISSN     \undefined \def \showISSN      #1{\unskip}     \fi
\ifx \showLCCN     \undefined \def \showLCCN      #1{\unskip}     \fi
\ifx \shownote     \undefined \def \shownote      #1{#1}          \fi
\ifx \showarticletitle \undefined \def \showarticletitle #1{#1}   \fi
\ifx \showURL      \undefined \def \showURL       {\relax}        \fi
\providecommand\bibfield[2]{#2}
\providecommand\bibinfo[2]{#2}
\providecommand\natexlab[1]{#1}
\providecommand\showeprint[2][]{arXiv:#2}

\bibitem[Mem(2018)]%
        {Memcached-link}
 \bibinfo{year}{2018}\natexlab{}.
\newblock \bibinfo{booktitle}{\emph{Memcached}}.
\newblock
\urldef\tempurl%
\url{https://memcached.org/}
\showURL{%
\tempurl}


\bibitem[sil(2018)]%
        {silesia}
 \bibinfo{year}{2018}\natexlab{}.
\newblock \bibinfo{title}{Silesia Corpus}.
\newblock \bibinfo{howpublished}{\url{https://github.com/MiloszKrajewski/SilesiaCorpus}}.
\newblock


\bibitem[xca(2022)]%
        {xcache-link}
 \bibinfo{year}{2022}\natexlab{}.
\newblock \bibinfo{title}{xcache}.
\newblock \bibinfo{howpublished}{\url{https://github.com/XimalayaCloud/xcache}}.
\newblock


\bibitem[BTC(2023)]%
        {BTC-link}
 \bibinfo{year}{2023}\natexlab{}.
\newblock \bibinfo{booktitle}{\emph{{Bitstamp Exchange Data}}}.
\newblock
\newblock
\shownote{\url{https://www.cryptodatadownload.com/data/bitstamp/}}.


\bibitem[FIO(2023)]%
        {FIO-link}
 \bibinfo{year}{2023}\natexlab{}.
\newblock \bibinfo{booktitle}{\emph{{Flexible I/O Tester}}}.
\newblock
\newblock
\shownote{\url{https://github.com/axboe/fio}}.


\bibitem[Red(2023)]%
        {Redis-link}
 \bibinfo{year}{2023}\natexlab{}.
\newblock \bibinfo{booktitle}{\emph{Redis}}.
\newblock
\urldef\tempurl%
\url{https://redis.io}
\showURL{%
\tempurl}


\bibitem[Roc(2023)]%
        {RocksDB-link}
 \bibinfo{year}{2023}\natexlab{}.
\newblock \bibinfo{booktitle}{\emph{RocksDB}}.
\newblock
\urldef\tempurl%
\url{https://rocksdb.org}
\showURL{%
\tempurl}


\bibitem[Cac(2024)]%
        {CacheLib-link}
 \bibinfo{year}{2024}\natexlab{}.
\newblock \bibinfo{booktitle}{\emph{CacheLib}}.
\newblock
\urldef\tempurl%
\url{https://github.com/facebook/CacheLib}
\showURL{%
\tempurl}


\bibitem[pik(2024)]%
        {pika-link}
 \bibinfo{year}{2024}\natexlab{}.
\newblock \bibinfo{title}{Pika}.
\newblock \bibinfo{howpublished}{\url{https://github.com/OpenAtomFoundation/pika}}.
\newblock


\bibitem[Sma(2024)]%
        {SmartSSD-link}
 \bibinfo{year}{2024}\natexlab{}.
\newblock \bibinfo{booktitle}{\emph{{Samsung SmartSSD}}}.
\newblock
\urldef\tempurl%
\url{https://semiconductor.samsung.com/ssd/smart-ssd/}
\showURL{%
\tempurl}


\bibitem[Sca(2024)]%
        {ScaleFlux-link}
 \bibinfo{year}{2024}\natexlab{}.
\newblock \bibinfo{booktitle}{\emph{ScaleFlux Computational Storage}}.
\newblock
\urldef\tempurl%
\url{http://scaleflux.com}
\showURL{%
\tempurl}


\bibitem[Wir(2024)]%
        {WiredTiger-link}
 \bibinfo{year}{2024}\natexlab{}.
\newblock \bibinfo{booktitle}{\emph{WiredTiger}}.
\newblock
\urldef\tempurl%
\url{https://github.com/wiredtiger/}
\showURL{%
\tempurl}


\bibitem[Zsw(2024)]%
        {Zswap-link}
 \bibinfo{year}{2024}\natexlab{}.
\newblock \bibinfo{booktitle}{\emph{Zswp}}.
\newblock
\newblock
\shownote{\url{https://wiki.archlinux.org/title/Zswap}}.


\bibitem[Adams et~al\mbox{.}(2020)]%
        {adams2020monarch}
\bibfield{author}{\bibinfo{person}{Colin Adams}, \bibinfo{person}{Luis Alonso}, \bibinfo{person}{Benjamin Atkin}, \bibinfo{person}{John Banning}, \bibinfo{person}{Sumeer Bhola}, \bibinfo{person}{Rick Buskens}, \bibinfo{person}{Ming Chen}, \bibinfo{person}{Xi Chen}, \bibinfo{person}{Yoo Chung}, \bibinfo{person}{Qin Jia}, {et~al\mbox{.}}} \bibinfo{year}{2020}\natexlab{}.
\newblock \showarticletitle{Monarch: Google's planet-scale in-memory time series database}.
\newblock \bibinfo{journal}{\emph{Proceedings of the VLDB Endowment}} \bibinfo{volume}{13}, \bibinfo{number}{12} (\bibinfo{year}{2020}), \bibinfo{pages}{3181--3194}.
\newblock


\bibitem[Ao(2023)]%
        {Ao23}
\bibfield{author}{\bibinfo{person}{Bryan Ao}.} \bibinfo{year}{2023}\natexlab{}.
\newblock \bibinfo{booktitle}{\emph{NAND Flash Prices Expected to Stabilize and Rebound in Q4, Projected to Remain Steady or Increase 0-5\%, Says TrendForce}}.
\newblock
\newblock
\shownote{\url{https://www.trendforce.com/presscenter/news/20230911-11839.html}}.


\bibitem[Atikoglu et~al\mbox{.}(2012)]%
        {Atikoglu12}
\bibfield{author}{\bibinfo{person}{Berk Atikoglu}, \bibinfo{person}{Yuehai Xu}, \bibinfo{person}{Eitan Frachtenberg}, \bibinfo{person}{Song Jiang}, {and} \bibinfo{person}{Mike Paleczny}.} \bibinfo{year}{2012}\natexlab{}.
\newblock \showarticletitle{{Workload Analysis of a Large-scale Key-Value Store}}. In \bibinfo{booktitle}{\emph{Proceedings of ACM SIGMETRICS Conference on Measurement and Modeling of Computer Systems (SIGMETRICS)}}. \bibinfo{pages}{53--64}.
\newblock


\bibitem[Balmau et~al\mbox{.}(2017)]%
        {balmau2017triad}
\bibfield{author}{\bibinfo{person}{Oana Balmau}, \bibinfo{person}{Diego Didona}, \bibinfo{person}{Rachid Guerraoui}, \bibinfo{person}{Willy Zwaenepoel}, \bibinfo{person}{Huapeng Yuan}, \bibinfo{person}{Aashray Arora}, \bibinfo{person}{Karan Gupta}, {and} \bibinfo{person}{Pavan Konka}.} \bibinfo{year}{2017}\natexlab{}.
\newblock \showarticletitle{{TRIAD}: Creating Synergies Between Memory, Disk and Log in Log Structured Key-Value Stores}. In \bibinfo{booktitle}{\emph{Proceedings of USENIX Annual Technical Conference (ATC)}}. \bibinfo{pages}{363--375}.
\newblock


\bibitem[Barbalace and Do(2021)]%
        {barbalacecomputational-21}
\bibfield{author}{\bibinfo{person}{Antonio Barbalace} {and} \bibinfo{person}{Jaeyoung Do}.} \bibinfo{year}{2021}\natexlab{}.
\newblock \showarticletitle{Computational Storage: Where Are We Today?}. In \bibinfo{booktitle}{\emph{Proc. of Annual Conference on Innovative Data Systems Research (CIDR)}}.
\newblock


\bibitem[Berg et~al\mbox{.}(2020)]%
        {berg2020cachelib}
\bibfield{author}{\bibinfo{person}{Ben Berg}, \bibinfo{person}{Daniel Berger}, \bibinfo{person}{Sara McAllister}, \bibinfo{person}{Isaac Grosof}, \bibinfo{person}{Sathya Gunasekar}, \bibinfo{person}{Jimmy Lu}, \bibinfo{person}{Michael Uhlar}, \bibinfo{person}{Jim Carrig}, \bibinfo{person}{Nathan Beckmann}, \bibinfo{person}{Mor Harchol-Balter}, {et~al\mbox{.}}} \bibinfo{year}{2020}\natexlab{}.
\newblock \showarticletitle{The CacheLib caching engine: Design and experiences at scale}. In \bibinfo{booktitle}{\emph{USENIX Symposium on Operating Systems Design and Implementation (OSDI)}}.
\newblock


\bibitem[Cao et~al\mbox{.}(2020)]%
        {cao2020polardb}
\bibfield{author}{\bibinfo{person}{Wei Cao}, \bibinfo{person}{Yang Liu}, \bibinfo{person}{Zhushi Cheng}, \bibinfo{person}{Ning Zheng}, \bibinfo{person}{Wei Li}, \bibinfo{person}{Wenjie Wu}, \bibinfo{person}{Linqiang Ouyang}, \bibinfo{person}{Peng Wang}, \bibinfo{person}{Yijing Wang}, \bibinfo{person}{Ray Kuan}, \bibinfo{person}{Zhenjun Liu}, \bibinfo{person}{Feng Zhu}, {and} \bibinfo{person}{Tong Zhang}.} \bibinfo{year}{2020}\natexlab{}.
\newblock \showarticletitle{{POLARDB meets computational storage: Efficiently support analytical workloads in cloud-native relational database}}. In \bibinfo{booktitle}{\emph{USENIX Conference on File and Storage Technologies (FAST)}}. \bibinfo{pages}{29--41}.
\newblock


\bibitem[Chandramouli et~al\mbox{.}(2018)]%
        {chandramouli2018faster}
\bibfield{author}{\bibinfo{person}{Badrish Chandramouli}, \bibinfo{person}{Guna Prasaad}, \bibinfo{person}{Donald Kossmann}, \bibinfo{person}{Justin Levandoski}, \bibinfo{person}{James Hunter}, {and} \bibinfo{person}{Mike Barnett}.} \bibinfo{year}{2018}\natexlab{}.
\newblock \showarticletitle{Faster: A concurrent key-value store with in-place updates}. In \bibinfo{booktitle}{\emph{Proceedings of the International Conference on Management of Data (SIGMOD)}}. \bibinfo{pages}{275--290}.
\newblock
\urldef\tempurl%
\url{https://doi.org/10.1145/3183713.3196898}
\showDOI{\tempurl}


\bibitem[Chen et~al\mbox{.}(2021a)]%
        {chen2021spandb}
\bibfield{author}{\bibinfo{person}{Hao Chen}, \bibinfo{person}{Chaoyi Ruan}, \bibinfo{person}{Cheng Li}, \bibinfo{person}{Xiaosong Ma}, {and} \bibinfo{person}{Yinlong Xu}.} \bibinfo{year}{2021}\natexlab{a}.
\newblock \showarticletitle{SpanDB: A fast, cost-effective LSM-tree based KV store on hybrid storage}. In \bibinfo{booktitle}{\emph{USENIX Conference on File and Storage Technologies (FAST)}}. \bibinfo{pages}{17--32}.
\newblock


\bibitem[Chen et~al\mbox{.}(2020)]%
        {chen2020hotring}
\bibfield{author}{\bibinfo{person}{Jiqiang Chen}, \bibinfo{person}{Liang Chen}, \bibinfo{person}{Sheng Wang}, \bibinfo{person}{Guoyun Zhu}, \bibinfo{person}{Yuanyuan Sun}, \bibinfo{person}{Huan Liu}, {and} \bibinfo{person}{Feifei Li}.} \bibinfo{year}{2020}\natexlab{}.
\newblock \showarticletitle{{HotRing}: A Hotspot-Aware In-Memory Key-Value Store}. In \bibinfo{booktitle}{\emph{USENIX Conference on File and Storage Technologies (FAST)}}. \bibinfo{pages}{239--252}.
\newblock
\showISBNx{9781939133120}


\bibitem[Chen et~al\mbox{.}(2021b)]%
        {chen2021kallaxdb}
\bibfield{author}{\bibinfo{person}{Xubin Chen}, \bibinfo{person}{Ning Zheng}, \bibinfo{person}{Shukun Xu}, \bibinfo{person}{Yifan Qiao}, \bibinfo{person}{Yang Liu}, \bibinfo{person}{Jiangpeng Li}, {and} \bibinfo{person}{Tong Zhang}.} \bibinfo{year}{2021}\natexlab{b}.
\newblock \showarticletitle{{KallaxDB}: A Table-less Hash-based Key-Value Store on Storage Hardware with Built-in Transparent Compression}. In \bibinfo{booktitle}{\emph{Proceedings of the International Workshop on Data Management on New Hardware (DaMoN)}}. \bibinfo{pages}{1--10}.
\newblock


\bibitem[Choukse et~al\mbox{.}(2018)]%
        {choukse2018compresso}
\bibfield{author}{\bibinfo{person}{Esha Choukse}, \bibinfo{person}{Mattan Erez}, {and} \bibinfo{person}{Alaa~R Alameldeen}.} \bibinfo{year}{2018}\natexlab{}.
\newblock \showarticletitle{Compresso: Pragmatic main memory compression}. In \bibinfo{booktitle}{\emph{2018 51st Annual IEEE/ACM International Symposium on Microarchitecture (MICRO)}}. IEEE, \bibinfo{pages}{546--558}.
\newblock


\bibitem[Dageville et~al\mbox{.}(2016)]%
        {dageville2016snowflake}
\bibfield{author}{\bibinfo{person}{Benoit Dageville}, \bibinfo{person}{Thierry Cruanes}, \bibinfo{person}{Marcin Zukowski}, \bibinfo{person}{Vadim Antonov}, \bibinfo{person}{Artin Avanes}, \bibinfo{person}{Jon Bock}, \bibinfo{person}{Jonathan Claybaugh}, \bibinfo{person}{Daniel Engovatov}, \bibinfo{person}{Martin Hentschel}, \bibinfo{person}{Jiansheng Huang}, {et~al\mbox{.}}} \bibinfo{year}{2016}\natexlab{}.
\newblock \showarticletitle{The snowflake elastic data warehouse}. In \bibinfo{booktitle}{\emph{Proceedings of the International Conference on Management of Data (SIGMOD)}}. \bibinfo{pages}{215--226}.
\newblock


\bibitem[Dayan and Idreos(2018)]%
        {dayan2018dostoevsky}
\bibfield{author}{\bibinfo{person}{Niv Dayan} {and} \bibinfo{person}{Stratos Idreos}.} \bibinfo{year}{2018}\natexlab{}.
\newblock \showarticletitle{{Dostoevsky: Better space-time trade-offs for LSM-tree based key-value stores via adaptive removal of superfluous merging}}. In \bibinfo{booktitle}{\emph{Proceedings of the International Conference on Management of Data (SIGMOD)}}. ACM, \bibinfo{pages}{505--520}.
\newblock


\bibitem[Ding et~al\mbox{.}(2022)]%
        {ding2022trianglekv}
\bibfield{author}{\bibinfo{person}{Chen Ding}, \bibinfo{person}{Ting Yao}, \bibinfo{person}{Hong Jiang}, \bibinfo{person}{Qiu Cui}, \bibinfo{person}{Liu Tang}, \bibinfo{person}{Yiwen Zhang}, \bibinfo{person}{Jiguang Wan}, {and} \bibinfo{person}{Zhihu Tan}.} \bibinfo{year}{2022}\natexlab{}.
\newblock \showarticletitle{{TriangleKV: Reducing write stalls and write amplification in LSM-tree based KV stores with triangle container in NVM}}.
\newblock \bibinfo{journal}{\emph{IEEE Transactions on Parallel and Distributed Systems}} \bibinfo{volume}{33}, \bibinfo{number}{12} (\bibinfo{year}{2022}), \bibinfo{pages}{4339--4352}.
\newblock


\bibitem[Dong et~al\mbox{.}(2021)]%
        {dong2021rocksdb}
\bibfield{author}{\bibinfo{person}{Siying Dong}, \bibinfo{person}{Andrew Kryczka}, \bibinfo{person}{Yanqin Jin}, {and} \bibinfo{person}{Michael Stumm}.} \bibinfo{year}{2021}\natexlab{}.
\newblock \showarticletitle{Rocksdb: Evolution of development priorities in a key-value store serving large-scale applications}.
\newblock \bibinfo{journal}{\emph{ACM Transactions on Storage (TOS)}} \bibinfo{volume}{17}, \bibinfo{number}{4} (\bibinfo{year}{2021}), \bibinfo{pages}{1--32}.
\newblock


\bibitem[Duan et~al\mbox{.}(2023)]%
        {duan2023revisiting}
\bibfield{author}{\bibinfo{person}{Zhuohui Duan}, \bibinfo{person}{Jiabo Yao}, \bibinfo{person}{Haikun Liu}, \bibinfo{person}{Xiaofei Liao}, \bibinfo{person}{Hai Jin}, {and} \bibinfo{person}{Yu Zhang}.} \bibinfo{year}{2023}\natexlab{}.
\newblock \showarticletitle{Revisiting Log-Structured Merging for {KV} Stores in Hybrid Memory Systems}. In \bibinfo{booktitle}{\emph{ACM International Conference on Architectural Support for Programming Languages and Operating Systems (ASPLOS)}}. \bibinfo{pages}{674--687}.
\newblock


\bibitem[Ekman and Stenstrom(2005)]%
        {ekman2005robust}
\bibfield{author}{\bibinfo{person}{Magnus Ekman} {and} \bibinfo{person}{Per Stenstrom}.} \bibinfo{year}{2005}\natexlab{}.
\newblock \showarticletitle{A robust main-memory compression scheme}. In \bibinfo{booktitle}{\emph{32nd International Symposium on Computer Architecture (ISCA)}}. IEEE, \bibinfo{pages}{74--85}.
\newblock


\bibitem[Fan et~al\mbox{.}(2013)]%
        {fan2013memc3}
\bibfield{author}{\bibinfo{person}{Bin Fan}, \bibinfo{person}{David~G Andersen}, {and} \bibinfo{person}{Michael Kaminsky}.} \bibinfo{year}{2013}\natexlab{}.
\newblock \showarticletitle{MemC3: Compact and Concurrent MemCache with Dumber Caching and Smarter Hashing}. In \bibinfo{booktitle}{\emph{USENIX Symposium on Networked Systems Design and Implementation (NSDI)}}. \bibinfo{pages}{371--384}.
\newblock


\bibitem[F{\"a}rber et~al\mbox{.}(2012)]%
        {farber2012sap}
\bibfield{author}{\bibinfo{person}{Franz F{\"a}rber}, \bibinfo{person}{Sang~Kyun Cha}, \bibinfo{person}{J{\"u}rgen Primsch}, \bibinfo{person}{Christof Bornh{\"o}vd}, \bibinfo{person}{Stefan Sigg}, {and} \bibinfo{person}{Wolfgang Lehner}.} \bibinfo{year}{2012}\natexlab{}.
\newblock \showarticletitle{SAP HANA database: data management for modern business applications}.
\newblock \bibinfo{journal}{\emph{ACM Sigmod Record}} \bibinfo{volume}{40}, \bibinfo{number}{4} (\bibinfo{year}{2012}), \bibinfo{pages}{45--51}.
\newblock


\bibitem[Graefe et~al\mbox{.}(2011)]%
        {graefe2011modern}
\bibfield{author}{\bibinfo{person}{Goetz Graefe} {et~al\mbox{.}}} \bibinfo{year}{2011}\natexlab{}.
\newblock \showarticletitle{Modern B-tree techniques}.
\newblock \bibinfo{journal}{\emph{Foundations and Trends{\textregistered} in Databases}} \bibinfo{volume}{3}, \bibinfo{number}{4} (\bibinfo{year}{2011}), \bibinfo{pages}{203--402}.
\newblock


\bibitem[Hao et~al\mbox{.}({[n.\,d.]})]%
        {hao2021implementing}
\bibfield{author}{\bibinfo{person}{Jingpeng Hao}, \bibinfo{person}{Xubin Chen}, \bibinfo{person}{Yifan Qiao}, \bibinfo{person}{Yuyang Zhang}, {and} \bibinfo{person}{Tong Zhang}.} \bibinfo{year}{[n.\,d.]}\natexlab{}.
\newblock \showarticletitle{Implementing Flash-Cached Storage Systems Using Computational Storage Drive with Built-in Transparent Compression}. In \bibinfo{booktitle}{\emph{2021 IEEE International Conference on Networking, Architecture and Storage (NAS)}}. \bibinfo{pages}{1--8}.
\newblock


\bibitem[Huang et~al\mbox{.}(2019)]%
        {huang2019x}
\bibfield{author}{\bibinfo{person}{Gui Huang}, \bibinfo{person}{Xuntao Cheng}, \bibinfo{person}{Jianying Wang}, \bibinfo{person}{Yujie Wang}, \bibinfo{person}{Dengcheng He}, \bibinfo{person}{Tieying Zhang}, \bibinfo{person}{Feifei Li}, \bibinfo{person}{Sheng Wang}, \bibinfo{person}{Wei Cao}, {and} \bibinfo{person}{Qiang Li}.} \bibinfo{year}{2019}\natexlab{}.
\newblock \showarticletitle{{X-Engine: An optimized storage engine for large-scale E-commerce transaction processing}}. In \bibinfo{booktitle}{\emph{Proceedings of the International Conference on Management of Data (SIGMOD)}}. ACM, \bibinfo{pages}{651--665}.
\newblock


\bibitem[Huang et~al\mbox{.}(2023)]%
        {huang2023breathing}
\bibfield{author}{\bibinfo{person}{Kecheng Huang}, \bibinfo{person}{Zhaoyan Shen}, \bibinfo{person}{Zili Shao}, \bibinfo{person}{Tong Zhang}, {and} \bibinfo{person}{Feng Chen}.} \bibinfo{year}{2023}\natexlab{}.
\newblock \showarticletitle{Breathing New Life into an Old Tree: Resolving Logging Dilemma of {B+-tree} on Modern Computational Storage Drives}.
\newblock \bibinfo{journal}{\emph{Proceedings of the VLDB Endowment}} \bibinfo{volume}{17}, \bibinfo{number}{2} (\bibinfo{year}{2023}), \bibinfo{pages}{134--147}.
\newblock


\bibitem[Kwon et~al\mbox{.}(2021)]%
        {kwon2021fast}
\bibfield{author}{\bibinfo{person}{Dongup Kwon}, \bibinfo{person}{Dongryeong Kim}, \bibinfo{person}{Junehyuk Boo}, \bibinfo{person}{Wonsik Lee}, {and} \bibinfo{person}{Jangwoo Kim}.} \bibinfo{year}{2021}\natexlab{}.
\newblock \showarticletitle{A fast and flexible hardware-based virtualization mechanism for computational storage devices}. In \bibinfo{booktitle}{\emph{USENIX Annual Technical Conference (ATC)}}. \bibinfo{pages}{729--743}.
\newblock


\bibitem[Lagar-Cavilla et~al\mbox{.}(2019)]%
        {lagar2019software}
\bibfield{author}{\bibinfo{person}{Andres Lagar-Cavilla}, \bibinfo{person}{Junwhan Ahn}, \bibinfo{person}{Suleiman Souhlal}, \bibinfo{person}{Neha Agarwal}, \bibinfo{person}{Radoslaw Burny}, \bibinfo{person}{Shakeel Butt}, \bibinfo{person}{Jichuan Chang}, \bibinfo{person}{Ashwin Chaugule}, \bibinfo{person}{Nan Deng}, \bibinfo{person}{Junaid Shahid}, {et~al\mbox{.}}} \bibinfo{year}{2019}\natexlab{}.
\newblock \showarticletitle{Software-defined far memory in warehouse-scale computers}. In \bibinfo{booktitle}{\emph{Proceedings of the International Conference on Architectural Support for Programming Languages and Operating Systems}}. \bibinfo{pages}{317--330}.
\newblock


\bibitem[Lahiri et~al\mbox{.}(2015)]%
        {lahiri2015oracle}
\bibfield{author}{\bibinfo{person}{Tirthankar Lahiri}, \bibinfo{person}{Shasank Chavan}, \bibinfo{person}{Maria Colgan}, \bibinfo{person}{Dinesh Das}, \bibinfo{person}{Amit Ganesh}, \bibinfo{person}{Mike Gleeson}, \bibinfo{person}{Sanket Hase}, \bibinfo{person}{Allison Holloway}, \bibinfo{person}{Jesse Kamp}, \bibinfo{person}{Teck-Hua Lee}, {et~al\mbox{.}}} \bibinfo{year}{2015}\natexlab{}.
\newblock \showarticletitle{Oracle database in-memory: A dual format in-memory database}. In \bibinfo{booktitle}{\emph{IEEE International Conference on Data Engineering (ICDE)}}. \bibinfo{pages}{1253--1258}.
\newblock


\bibitem[Levandoski et~al\mbox{.}(2013)]%
        {bwtree-13}
\bibfield{author}{\bibinfo{person}{Justin~J. Levandoski}, \bibinfo{person}{David~B. Lomet}, {and} \bibinfo{person}{Sudipta Sengupta}.} \bibinfo{year}{2013}\natexlab{}.
\newblock \showarticletitle{{The Bw-Tree: {A} B-tree for new hardware platforms}}. In \bibinfo{booktitle}{\emph{IEEE International Conference on Data Engineering (ICDE)}}. \bibinfo{pages}{302--313}.
\newblock


\bibitem[Li et~al\mbox{.}({[n.\,d.]})]%
        {li2014nitro}
\bibfield{author}{\bibinfo{person}{Cheng Li}, \bibinfo{person}{Philip Shilane}, \bibinfo{person}{Fred Douglis}, \bibinfo{person}{Hyong Shim}, \bibinfo{person}{Stephen Smaldone}, {and} \bibinfo{person}{Grant Wallace}.} \bibinfo{year}{[n.\,d.]}\natexlab{}.
\newblock \showarticletitle{Nitro: A Capacity-Optimized SSD Cache for Primary Storage}. In \bibinfo{booktitle}{\emph{2014 USENIX Annual Technical Conference (USENIX ATC)}}. \bibinfo{pages}{501--512}.
\newblock


\bibitem[Li et~al\mbox{.}(2016)]%
        {li2016cachededup}
\bibfield{author}{\bibinfo{person}{Wenji Li}, \bibinfo{person}{Gregory Jean-Baptise}, \bibinfo{person}{Juan Riveros}, \bibinfo{person}{Giri Narasimhan}, \bibinfo{person}{Tony Zhang}, {and} \bibinfo{person}{Ming Zhao}.} \bibinfo{year}{2016}\natexlab{}.
\newblock \showarticletitle{CacheDedup: In-line Deduplication for Flash Caching}. In \bibinfo{booktitle}{\emph{14th USENIX Conference on File and Storage Technologies (FAST)}}. \bibinfo{pages}{301--314}.
\newblock


\bibitem[Lim et~al\mbox{.}(2014)]%
        {lim2014mica}
\bibfield{author}{\bibinfo{person}{Hyeontaek Lim}, \bibinfo{person}{Dongsu Han}, \bibinfo{person}{David~G Andersen}, {and} \bibinfo{person}{Michael Kaminsky}.} \bibinfo{year}{2014}\natexlab{}.
\newblock \showarticletitle{{MICA}: A Holistic Approach to Fast In-Memory Key-Value Storage}. In \bibinfo{booktitle}{\emph{USENIX Symposium on Networked Systems Design and Implementation (NSDI)}}. \bibinfo{pages}{429--444}.
\newblock


\bibitem[Mao et~al\mbox{.}(2012)]%
        {mao2012cache}
\bibfield{author}{\bibinfo{person}{Yandong Mao}, \bibinfo{person}{Eddie Kohler}, {and} \bibinfo{person}{Robert~Tappan Morris}.} \bibinfo{year}{2012}\natexlab{}.
\newblock \showarticletitle{Cache craftiness for fast multicore key-value storage}. In \bibinfo{booktitle}{\emph{Proceedings of the european conference on Computer Systems}}. \bibinfo{pages}{183--196}.
\newblock


\bibitem[Maruf et~al\mbox{.}(2023)]%
        {Maruf_2023}
\bibfield{author}{\bibinfo{person}{Hasan~Al Maruf}, \bibinfo{person}{Hao Wang}, \bibinfo{person}{Abhishek Dhanotia}, \bibinfo{person}{Johannes Weiner}, \bibinfo{person}{Niket Agarwal}, \bibinfo{person}{Pallab Bhattacharya}, \bibinfo{person}{Chris Petersen}, \bibinfo{person}{Mosharaf Chowdhury}, \bibinfo{person}{Shobhit Kanaujia}, {and} \bibinfo{person}{Prakash Chauhan}.} \bibinfo{year}{2023}\natexlab{}.
\newblock \showarticletitle{TPP: Transparent Page Placement for CXL-Enabled Tiered-Memory}. In \bibinfo{booktitle}{\emph{Proceedings of the International Conference on Architectural Support for Programming Languages and Operating Systems, Volume 3}} \emph{(\bibinfo{series}{ASPLOS})}.
\newblock
\urldef\tempurl%
\url{https://doi.org/10.1145/3582016.3582063}
\showDOI{\tempurl}


\bibitem[McAllister et~al\mbox{.}(2021)]%
        {mcallister2021kangaroo}
\bibfield{author}{\bibinfo{person}{Sara McAllister}, \bibinfo{person}{Benjamin Berg}, \bibinfo{person}{Julian Tutuncu-Macias}, \bibinfo{person}{Juncheng Yang}, \bibinfo{person}{Sathya Gunasekar}, \bibinfo{person}{Jimmy Lu}, \bibinfo{person}{Daniel~S Berger}, \bibinfo{person}{Nathan Beckmann}, {and} \bibinfo{person}{Gregory~R Ganger}.} \bibinfo{year}{2021}\natexlab{}.
\newblock \showarticletitle{Kangaroo: Caching billions of tiny objects on flash}. In \bibinfo{booktitle}{\emph{Proceedings of the ACM Symposium on Operating Systems Principles (SOSP)}}. \bibinfo{pages}{243--262}.
\newblock


\bibitem[Morgan(2020)]%
        {Morgan20}
\bibfield{author}{\bibinfo{person}{Timothy~Prickett Morgan}.} \bibinfo{year}{2020}\natexlab{}.
\newblock \bibinfo{booktitle}{\emph{CXL and Gen-Z Iron Out A Coherent Interconnect Strategy}}.
\newblock
\newblock
\shownote{\url{https://www.nextplatform.com/2020/04/03/cxl-and-gen-z-iron-out-a-coherent-interconnect-strategy/}}.


\bibitem[Olšan(2023)]%
        {Olsan23}
\bibfield{author}{\bibinfo{person}{Jan Olšan}.} \bibinfo{year}{2023}\natexlab{}.
\newblock \bibinfo{booktitle}{\emph{The days of SSDs getting cheaper are over. Prices are starting to rise}}.
\newblock
\newblock
\shownote{\url{https://www.hwcooling.net/en/the-days-of-ssds-getting-cheaper-are-over-prices-will-rise/}}.


\bibitem[Ousterhout et~al\mbox{.}(2015)]%
        {ousterhout2015ramcloud}
\bibfield{author}{\bibinfo{person}{John Ousterhout}, \bibinfo{person}{Arjun Gopalan}, \bibinfo{person}{Ashish Gupta}, \bibinfo{person}{Ankita Kejriwal}, \bibinfo{person}{Collin Lee}, \bibinfo{person}{Behnam Montazeri}, \bibinfo{person}{Diego Ongaro}, \bibinfo{person}{Seo~Jin Park}, \bibinfo{person}{Henry Qin}, \bibinfo{person}{Mendel Rosenblum}, {et~al\mbox{.}}} \bibinfo{year}{2015}\natexlab{}.
\newblock \showarticletitle{The RAMCloud storage system}.
\newblock \bibinfo{journal}{\emph{ACM Transactions on Computer Systems (TOCS)}} \bibinfo{volume}{33}, \bibinfo{number}{3} (\bibinfo{year}{2015}), \bibinfo{pages}{1--55}.
\newblock


\bibitem[O’Neil et~al\mbox{.}(1996)]%
        {o1996log}
\bibfield{author}{\bibinfo{person}{Patrick O’Neil}, \bibinfo{person}{Edward Cheng}, \bibinfo{person}{Dieter Gawlick}, {and} \bibinfo{person}{Elizabeth O’Neil}.} \bibinfo{year}{1996}\natexlab{}.
\newblock \showarticletitle{The log-structured merge-tree (LSM-tree)}.
\newblock \bibinfo{journal}{\emph{Acta Informatica}}  \bibinfo{volume}{33} (\bibinfo{year}{1996}), \bibinfo{pages}{351--385}.
\newblock


\bibitem[Pekhimenko et~al\mbox{.}(2013)]%
        {pekhimenko2013linearly}
\bibfield{author}{\bibinfo{person}{Gennady Pekhimenko}, \bibinfo{person}{Vivek Seshadri}, \bibinfo{person}{Yoongu Kim}, \bibinfo{person}{Hongyi Xin}, \bibinfo{person}{Onur Mutlu}, \bibinfo{person}{Phillip~B Gibbons}, \bibinfo{person}{Michael~A Kozuch}, {and} \bibinfo{person}{Todd~C Mowry}.} \bibinfo{year}{2013}\natexlab{}.
\newblock \showarticletitle{Linearly compressed pages: A low-complexity, low-latency main memory compression framework}. In \bibinfo{booktitle}{\emph{Proceedings of the Annual IEEE/ACM International Symposium on Microarchitecture}}. \bibinfo{pages}{172--184}.
\newblock


\bibitem[Pelkonen et~al\mbox{.}(2015)]%
        {pelkonen2015gorilla}
\bibfield{author}{\bibinfo{person}{Tuomas Pelkonen}, \bibinfo{person}{Scott Franklin}, \bibinfo{person}{Justin Teller}, \bibinfo{person}{Paul Cavallaro}, \bibinfo{person}{Qi Huang}, \bibinfo{person}{Justin Meza}, {and} \bibinfo{person}{Kaushik Veeraraghavan}.} \bibinfo{year}{2015}\natexlab{}.
\newblock \showarticletitle{Gorilla: A fast, scalable, in-memory time series database}.
\newblock \bibinfo{journal}{\emph{Proceedings of the VLDB Endowment}} \bibinfo{volume}{8}, \bibinfo{number}{12} (\bibinfo{year}{2015}), \bibinfo{pages}{1816--1827}.
\newblock


\bibitem[Raju et~al\mbox{.}(2017)]%
        {PebblesDB-17}
\bibfield{author}{\bibinfo{person}{Pandian Raju}, \bibinfo{person}{Rohan Kadekodi}, \bibinfo{person}{Vijay Chidambaram}, {and} \bibinfo{person}{Ittai Abraham}.} \bibinfo{year}{2017}\natexlab{}.
\newblock \showarticletitle{{PebblesDB}: Building Key-Value Stores Using Fragmented Log-Structured Merge Trees}. In \bibinfo{booktitle}{\emph{Proceedings of the Symposium on Operating Systems Principles (SOSP)}}. \bibinfo{pages}{497--514}.
\newblock


\bibitem[Vin{\c{c}}on et~al\mbox{.}(2022)]%
        {vinccon2022near}
\bibfield{author}{\bibinfo{person}{Tobias Vin{\c{c}}on}, \bibinfo{person}{Christian Kn{\"o}dler}, \bibinfo{person}{Leonardo Solis-Vasquez}, \bibinfo{person}{Arthur Bernhardt}, \bibinfo{person}{Sajjad Tamimi}, \bibinfo{person}{Lukas Weber}, \bibinfo{person}{Florian Stock}, \bibinfo{person}{Andreas Koch}, {and} \bibinfo{person}{Ilia Petrov}.} \bibinfo{year}{2022}\natexlab{}.
\newblock \showarticletitle{Near-data processing in database systems on native computational storage under {HTAP} workloads}.
\newblock \bibinfo{journal}{\emph{Proceedings of the VLDB Endowment}} \bibinfo{volume}{15}, \bibinfo{number}{10} (\bibinfo{year}{2022}), \bibinfo{pages}{1991--2004}.
\newblock


\bibitem[Wang et~al\mbox{.}({[n.\,d.]})]%
        {wang2020austere}
\bibfield{author}{\bibinfo{person}{Qiuping Wang}, \bibinfo{person}{Jinhong Li}, \bibinfo{person}{Wen Xia}, \bibinfo{person}{Erik Kruus}, \bibinfo{person}{Biplob Debnath}, {and} \bibinfo{person}{Patrick~PC Lee}.} \bibinfo{year}{[n.\,d.]}\natexlab{}.
\newblock \showarticletitle{Austere Flash Caching with Deduplication and Compression}. In \bibinfo{booktitle}{\emph{2020 USENIX Annual Technical Conference (USENIX ATC)}}. \bibinfo{pages}{713--726}.
\newblock


\bibitem[Weiner et~al\mbox{.}(2022)]%
        {weiner2022tmo}
\bibfield{author}{\bibinfo{person}{Johannes Weiner}, \bibinfo{person}{Niket Agarwal}, \bibinfo{person}{Dan Schatzberg}, \bibinfo{person}{Leon Yang}, \bibinfo{person}{Hao Wang}, \bibinfo{person}{Blaise Sanouillet}, \bibinfo{person}{Bikash Sharma}, \bibinfo{person}{Tejun Heo}, \bibinfo{person}{Mayank Jain}, \bibinfo{person}{Chunqiang Tang}, {et~al\mbox{.}}} \bibinfo{year}{2022}\natexlab{}.
\newblock \showarticletitle{Tmo: transparent memory offloading in datacenters}. In \bibinfo{booktitle}{\emph{Proceedings of the International Conference on Architectural Support for Programming Languages and Operating Systems}}. \bibinfo{pages}{609--621}.
\newblock


\bibitem[White(2023)]%
        {White23}
\bibfield{author}{\bibinfo{person}{Monica~J. White}.} \bibinfo{year}{2023}\natexlab{}.
\newblock \bibinfo{booktitle}{\emph{The era of cheap SSDs is about to end}}.
\newblock
\newblock
\shownote{\url{https://www.digitaltrends.com/computing/samsung-flash-nand-chips-price-increase/}}.


\bibitem[Yao et~al\mbox{.}(2020)]%
        {yao2020matrixkv}
\bibfield{author}{\bibinfo{person}{Ting Yao}, \bibinfo{person}{Yiwen Zhang}, \bibinfo{person}{Jiguang Wan}, \bibinfo{person}{Qiu Cui}, \bibinfo{person}{Liu Tang}, \bibinfo{person}{Hong Jiang}, \bibinfo{person}{Changsheng Xie}, {and} \bibinfo{person}{Xubin He}.} \bibinfo{year}{2020}\natexlab{}.
\newblock \showarticletitle{{MatrixKV}: Reducing Write Stalls and Write Amplification in {LSM}-tree Based {KV} Stores with Matrix Container in {NVM}}. In \bibinfo{booktitle}{\emph{USENIX Annual Technical Conference (ATC)}}. \bibinfo{pages}{17--31}.
\newblock


\bibitem[Zhan et~al\mbox{.}(2020)]%
        {zhan2020rangekv}
\bibfield{author}{\bibinfo{person}{Ling Zhan}, \bibinfo{person}{Kai Lu}, \bibinfo{person}{Zhilong Cheng}, {and} \bibinfo{person}{Jiguang Wan}.} \bibinfo{year}{2020}\natexlab{}.
\newblock \showarticletitle{{RangeKV: An efficient key-value store based on hybrid DRAM-NVM-SSD storage structure}}.
\newblock \bibinfo{journal}{\emph{IEEE Access}}  \bibinfo{volume}{8} (\bibinfo{year}{2020}), \bibinfo{pages}{154518--154529}.
\newblock


\bibitem[Zhang et~al\mbox{.}(2015)]%
        {zhang2015mega}
\bibfield{author}{\bibinfo{person}{Kai Zhang}, \bibinfo{person}{Kaibo Wang}, \bibinfo{person}{Yuan Yuan}, \bibinfo{person}{Lei Guo}, \bibinfo{person}{Rubao Lee}, {and} \bibinfo{person}{Xiaodong Zhang}.} \bibinfo{year}{2015}\natexlab{}.
\newblock \showarticletitle{Mega-kv: A case for gpus to maximize the throughput of in-memory key-value stores}.
\newblock \bibinfo{journal}{\emph{Proceedings of the VLDB Endowment}} \bibinfo{volume}{8}, \bibinfo{number}{11} (\bibinfo{year}{2015}), \bibinfo{pages}{1226--1237}.
\newblock


\bibitem[Zhao et~al\mbox{.}(2015)]%
        {zhao2015buri}
\bibfield{author}{\bibinfo{person}{Jishen Zhao}, \bibinfo{person}{Sheng Li}, \bibinfo{person}{Jichuan Chang}, \bibinfo{person}{John~L Byrne}, \bibinfo{person}{Laura~L Ramirez}, \bibinfo{person}{Kevin Lim}, \bibinfo{person}{Yuan Xie}, {and} \bibinfo{person}{Paolo Faraboschi}.} \bibinfo{year}{2015}\natexlab{}.
\newblock \showarticletitle{Buri: Scaling big-memory computing with hardware-based memory expansion}.
\newblock \bibinfo{journal}{\emph{ACM Transactions on Architecture and Code Optimization (TACO)}} \bibinfo{volume}{12}, \bibinfo{number}{3} (\bibinfo{year}{2015}), \bibinfo{pages}{1--24}.
\newblock


\bibitem[Zheng et~al\mbox{.}(2020)]%
        {Zheng-Hotstorage-20}
\bibfield{author}{\bibinfo{person}{Ning Zheng}, \bibinfo{person}{Xubin Chen}, \bibinfo{person}{Jiangpeng Li}, \bibinfo{person}{Qi Wu}, \bibinfo{person}{Yang Liu}, \bibinfo{person}{Yong Peng}, \bibinfo{person}{Fei Sun}, \bibinfo{person}{Hao Zhong}, {and} \bibinfo{person}{Tong Zhang}.} \bibinfo{year}{2020}\natexlab{}.
\newblock \showarticletitle{Re-think Data Management Software Design Upon the Arrival of Storage Hardware with Built-in Transparent Compression}. In \bibinfo{booktitle}{\emph{{USENIX} Workshop on Hot Topics in Storage and File Systems (HotStorage)}}.
\newblock


\bibitem[Ziv and Lempel(1977)]%
        {ziv1977universal}
\bibfield{author}{\bibinfo{person}{Jacob Ziv} {and} \bibinfo{person}{Abraham Lempel}.} \bibinfo{year}{1977}\natexlab{}.
\newblock \showarticletitle{A universal algorithm for sequential data compression}.
\newblock \bibinfo{journal}{\emph{IEEE Transactions on information theory}} \bibinfo{volume}{23}, \bibinfo{number}{3} (\bibinfo{year}{1977}), \bibinfo{pages}{337--343}.
\newblock
\urldef\tempurl%
\url{https://doi.org/10.1109/TIT.1977.1055714}
\showDOI{\tempurl}


\bibitem[Ziv and Lempel(1978)]%
        {ziv1978compression}
\bibfield{author}{\bibinfo{person}{Jacob Ziv} {and} \bibinfo{person}{Abraham Lempel}.} \bibinfo{year}{1978}\natexlab{}.
\newblock \showarticletitle{Compression of individual sequences via variable-rate coding}.
\newblock \bibinfo{journal}{\emph{IEEE transactions on Information Theory}} \bibinfo{volume}{24}, \bibinfo{number}{5} (\bibinfo{year}{1978}), \bibinfo{pages}{530--536}.
\newblock
\urldef\tempurl%
\url{https://doi.org/10.1109/TIT.1978.1055934}
\showDOI{\tempurl}


\end{thebibliography}

\end{document}